\documentclass{jfm}
\usepackage{graphicx}
\usepackage{epstopdf, epsfig}

\shorttitle{Asymmetry and Buoyancy Effects on Turbulent Multi-component Jets}
\shortauthor{M. Soleimani nia, B. Maxwell, P. Oshkai and N. Djilali}

\title{Measurements of Flow Velocity and Scalar Concentration in Turbulent Multi-component Jets: Asymmetry and Buoyancy Effects}

\author{Majid Soleimani nia\aff{1}
  \corresp{\email{majids@uvic.ca}},
  Brian Maxwell\aff{2},
  Peter Oshkai\aff{1}
 \and Ned Djilali\aff{1}}

\affiliation{\aff{1}Department of Mechanical Engineering and Institute for Integrated Energy Systems, University of Victoria, 
PO Box 1700 STN CSC, Victoria, BC, V8W 2Y2, Canada
\aff{2}Department of Mechanical and Aerospace Engineering, Case Western Reserve University,
10900 Euclid Avenue, Glennan 418, Cleveland Ohio, 44106, USA
}

\begin{document}

\maketitle

\begin{abstract}
Buoyancy effects and nozzle geometry can have a significant impact on turbulent jet dispersion. This work was motivated by applications involving hydrogen. Using helium as an experimental proxy, buoyant horizontal jets issuing from a round orifice on the side wall of a circular tube were analysed experimentally using particle image velocimetry (PIV) and planar laser-induced fluorescence (PLIF) techniques simultaneously to provide instantaneous and time-averaged flow fields of velocity and concentration. Effects of buoyancy and asymmetry on the resulting flow structure were studied over a range of Reynolds numbers and gas densities. Significant differences were found between the centreline trajectory, spreading rate, and velocity decay of conventional horizontal round axisymmetric jets issuing through flat plates and the pipeline leak-representative jets considered in the present study. The realistic pipeline jets were always asymmetric and found to deflect about the jet axis in the near field. In the far field, it was found that the realistic pipeline leak geometry causes buoyancy effects to dominate much sooner than expected compared to horizontal round jets issuing through flat plates. 
\end{abstract}


\section{Introduction}
Hydrogen, a carbon-free energy carrier, is currently viewed as a clean alternative to traditional hydrocarbon-based fuels for transportation and energy storage applications. It can burn or react with almost no pollution or green house gas emissions, and is commonly used in electrochemical fuel cells to power vehicle and electrical devices.  It is also used in an increasing number of power-to-gas systems to blend in the natural gas pipeline network. Despite these benefits, previous studies have shown that hydrogen jets resulting from an accidental leak are easily ignitable\citep{Veser2011IJoHE2351}, owing to a wide range of possible ignition limits (between 4\%-75\% by volume) \citep{Lewis1961}.	Therefore, modern safety standards for hydrogen storage infrastructure must be assured before widespread public use of hydrogen can become possible.  As a result, fundamental insight into the physics of hydrogen dispersion into ambient air from realistic flow geometries, such as small pipelines, is necessary to properly predict flow structures and flammability envelope associated with hydrogen outflow from accidental leaks.  Also, owing to the low molecular weight of hydrogen, buoyancy can significantly influence the development of the jet dispersion during a release scenario.  In the current investigation, we attempt to quantify the dispersion and release trajectories of horizontal buoyant jets experimentally, as they emerge from a realistic pipeline geometry, using state-of-the-art experimental imaging techniques.

In the last two decades, due to the rapid development of the hydrogen economy and use of hydrogen technologies, several experimental and numerical studies \citep{Chernyavsky2011IJoHE2645,Houf2008IJoHE1435,Schefer2008IJoHE4702,Schefer2008IJoHE6373,Xiao2011IJoHE2545} have investigated small-scale unintended hydrogen round jet release in ambient air, while others \citep{Ekoto2012IJoHE17446,Houf2013IJoHE8179,Houf2010IJoHE4758,Hajji2015IJoHE9747} studied different accidental hydrogen dispersion scenarios in enclosed and open spaces.  There has also been extensive work done to describe the evolution of axisymmetric round turbulent jets in terms of self-similarity correlations, obtained from statistical analysis from both experiments \citep{Lipari2011FTaC79,Ball2012PiAS1} and simulations \citep{Kaushik2015AJoFD1}.  In addition, some investigations \citep{L.K.2010JoFM59} have quantified the buoyancy effects on vertical round jets, while others \citep{Rodi1982,Carazzo2006JoFM137} have provided a quantitative study into the buoyancy effects on both turbulent buoyant/pure jets and plumes with analyzing of all available experimental data. Even though jets and plumes both have different states of partial or local self-similarity \citep{george1989self}, their global evolutions in the far field tends toward complete self-similarity through a universal route even in the presence of buoyancy. Large-scale structures of turbulence drive the evolution of the self-similarity profile, and buoyancy has an effect in exciting the coherent turbulent structures; this effect is more evident in the evolution of plumes into self-similarly much	sooner owing to buoyancy driven turbulence in the near field \citep{Carazzo2006JoFM137}. Horizontal buoyant jets, however, have been much less studied.  In general, increasing effects of buoyancy were found to correlate inversely to the Froude number in axisymmetry horizontal buoyant jets \citep{Ash2012}.

Previous measurements on axisymmetric round hydrogen jets \citep{Schefer2008IJoHE6373,Schefer2008IJoHE4702} revealed that, hydrogen jets show the same behavior to jets of helium \citep{Panchapakesan1993JoFM225}, propane and other hydrocarbon fuels \citep{Richards1993JoFM417}. In particular, the intensity of centreline velocity fluctuations are similar between the jet and plume regions. In contrast, mass fraction fluctuation intensities increased from a constant asymptotic value of about 0.23 in the jet region to 0.33-0.37 in the plume region \citep{Panchapakesan1993JoFM225,Schefer2008IJoHE4702}. It has also been well established that the mass fraction fluctuation intensities along the centerline and radial variations are also independent of initial density differences between ambient and jet fluids, and collapse onto the same curves, different curves in jet and plume regions, when plotted against the appropriate similarity variables \citep{Panchapakesan1993JoFM225,Schefer2008IJoHE4702,Schefer2008IJoHE6373,Pitts1991EiF125}.

It is noteworthy that all aforementioned studies, as well as related previous investigations on jets or plumes, have been limited to leaks through flat surfaces, where the direction of the jet mean flow was aligned with the flow origin. In reality, however, flow patterns and dispersion of accidental gas leaks would not be limited to flows through flat surfaces, and leaks through cracks in the side walls of circular pipes should also receive attention. To address this, a recent study was investigated for vertical buoyant jet evolutions through round holes from curved surfaces, numerically and experimentally \citep{Soleimaninia2018IJoHE,Soleimaninia2017,Maxwell2017}. Through this recent work, significant discrepancies were found between the evolution of axisymmetric round sharp-edged Orifice Plate (OP) jets through flat surfaces and those originating from curved surfaces.  Most notably, jet deflection from the vertical axis, and asymmetric dispersion patterns are always observed in realistic situations.  To our knowledge, however, the horizontal jet dispersion from curved surfaces has not yet been investigated.

To investigate the effects of asymmetry and buoyancy on the evolution of horizontal jets issuing from realistic pipeline geometries, jet release experiments were conducted with air and helium,  where flow patterns and dispersion of gas through a curved surface originating from a source whose original velocity components were nearly perpendicular to the direction of the ensuing jets. From now on, this jet configuration is referred as a 3D jet. A round hole as one of possible crack geometries, was considered in this study, although another possibility might include thin cracks around the tube or the faulty tube fittings \citep{Iverson2015IJoHE13134}, which is not considered here. The horizontal 3D jets were released through a 2 mm diameter round hole in the side of a round tube (closed at one end), with an outer diameter of 6.36 mm and 0.82 mm wall thickness. The outer-scale flow Reynolds numbers ($Re_{\delta}$), based on the orifice diameter, and Mach numbers ($Ma$) of the jets ranged from 19,000 to 51,500 and 0.4 to 1.2, respectively.  However, it is noted that for hydrogen jets of equivalent momentum flux, the expected Mach number and Reynolds number would be 1.5 and 55,915, respectively.  At these conditions, hydrogen is expected to behave very similar to the helium jets considered here \citep{Soleimaninia2018IJoHE}. These realistic jets were also compared to axisymmetric leaks through flat surfaces accordingly.  Particle imaging velocimetry (PIV) and planar laser-induced fluorescence (PLIF) were used to measure high-resolution instantaneous velocity and concentration fields, respectively.  The purpose of this investigation was to identify and characterize departures from standard axisymmetric jet conditions, and to highlight the buoyancy effect and asymmetric nature of the 3D jets, which ensued from a practical geometry arrangement. It should be noted that, the effect of pipe wall thickness of the crack geometry has not yet been investigated.  

\section{Experimental system and techniques}
\subsection{Flow facility}
\begin{figure}
	\centering
	a)\includegraphics[scale=0.25]{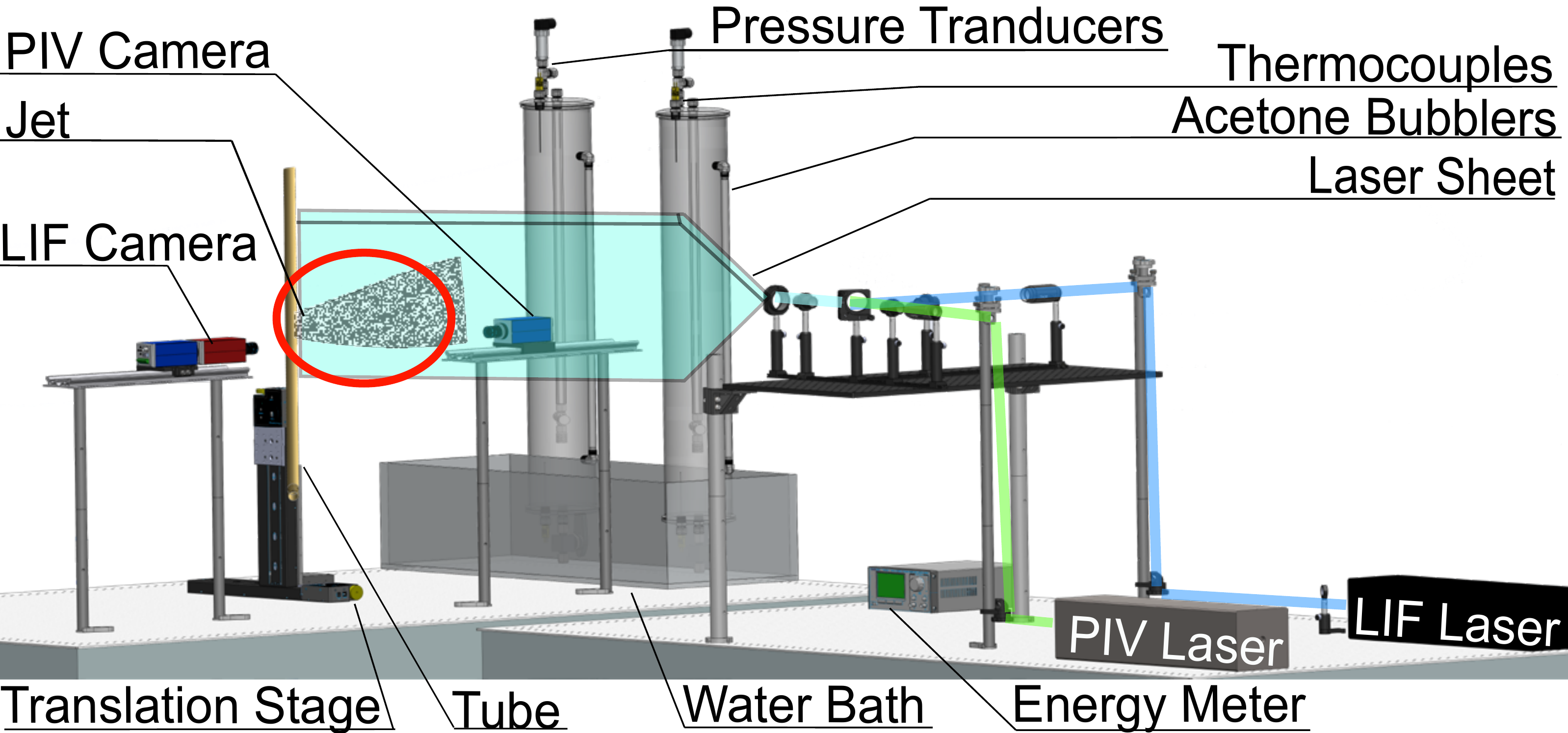}
	b)\includegraphics[scale=0.12]{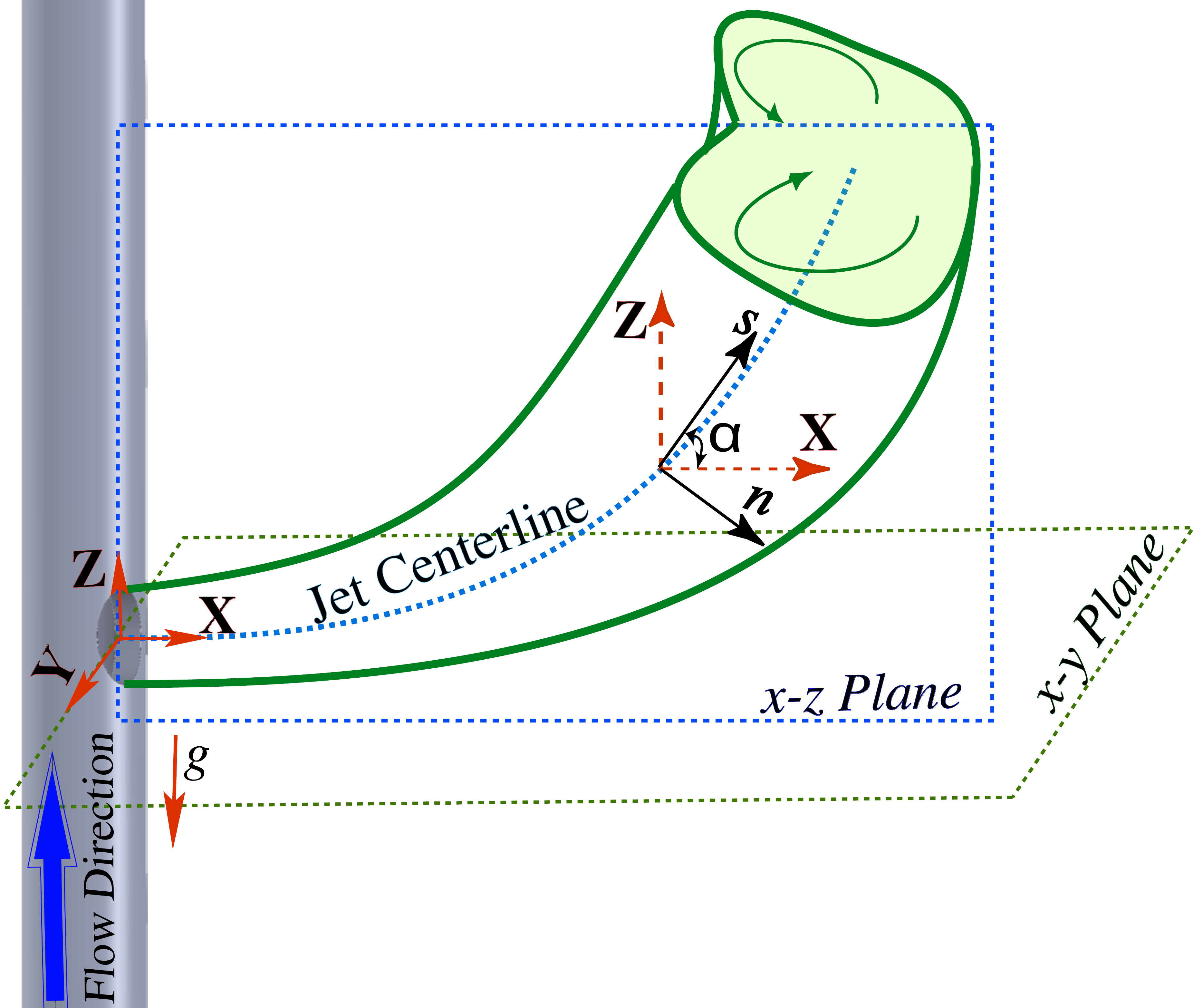}
	\caption{a)   {Schematic of the experimental layout.} b) Illustration of horizontal 3D jet flow measurement area   {(red inset in part a)}.}
	\label{fig.Experimental_Layout}
\end{figure}
Figure \ref{fig.Experimental_Layout}a, provides a schematic of the experimental setup used for this study. While, Figure \ref{fig.Experimental_Layout}b, {illustrates} the jet flow evolution from the tube orifice considered. To capture the three-dimensionality of the jet, measurements {were} obtained on two different  two-dimensional planes (denoted $x$-$z$ and $x$-$y$), as indicated, for both air and helium.  Also {shown} in the figure is the jet centreline, which acts as a reference from which measurements are later obtained in the $x$-$z$ plane.  Owing to potential deviation of the jet from the orifice axis ($x$-axis), the jet centreline tangent and normal lines are shown as $\textrm{\bf{s}}$ and $\textrm{\bf{n}}$ coordinates in the figure, respectively.

The experiments were conducted within a controlled stagnant environment, at room temperature and pressure (${T_o}\sim22^{\circ \textrm{C}}$, ${p_o}\sim100$ kPa). Flow controllers (Bronkhorst, EL-FLOW series) were used to control mass flow rates of dry air and pure scientific grade helium to the system, with a high accuracy (standard $\pm0.5\%$ of reading plus $\pm0.1\%$ full scale) and precision (within $0.2\% $ of the reading). Di-Ethyl-Hexyl-Sebacate (DEHS) tracer particles were used in PIV measurements, while acetone vapour used as fluorescent tracers for the PLIF experiments. After the test gas was mixed and seeded with the PIV and PLIF tracers, the flow entered the test section of the tube. Isothermal and isobaric conditions were ensured in all measurements. Further specific details can be found in \citep{Soleimaninia2018IJoHE}. The orifice, through which the gas dispersed, had a diameter of ${D}=2$ mm and was located sufficiently downstream along the tube length to ensure fully developed flow within the tube at the orifice location.  Within the tube, flow controllers were used to ensure fully developed subsonic and turbulent flow inside the tube.

\begin{table}
  \begin{center}
\def~{\hphantom{0}}
		{\small
			\begin{tabular}{lccccccccc}
				Jet & Orientation &  $Q$ & $\langle{\boldsymbol{u}_{j}}\rangle$ &$\rho_{{j}}$ &$\nu_{{j}}$  &${M} $  & $Ma$ &  $Fr$ & $Re_{\delta}$\\
				&, Type&  [L/min] & [m/s]  & [Kg/$\textrm{m}^{3}$] &  [$\textrm{m}^{2}$/s]  & [N/m]  &   & & \\[3pt]						
				Air&H, 3D&15&147.5 & 1.17 & $1.54 \times 10^{-5} $&50.9 & 0.43 & - &19,000 \\
				Air&V, 3D&15&147.5 & 1.17 & $1.54 \times 10^{-5} $&50.9 & 0.43 & - &19,000 \\
				Air&V, OP&15&127.6 & 1.17 & $1.54 \times 10^{-5} $&38.1 & 0.37 & - &16,500 \\
				He&H, 3D&35& 399.5& 0.165 &$1.21 \times 10^{-4} $  &51.3& 1.2 &$1.34\times10^{6} $&51,500 \\			
				He&V, 3D&35& 399.7& 0.165 &$1.21 \times 10^{-4} $  &51.4& 1.2 &$1.34\times10^{6} $&51,500 \\			
				He&V, OP&35& 341.9& 0.165 &$1.21 \times 10^{-4} $  &38.3& 1 &$9.8\times10^{5} $&44,200 \\			
		\end{tabular}}
		\caption{Flow properties}
		\label{tab.Flow properties_HSlot1}
	\end{center}
\end{table}

In order to compare the behaviour of both test gases, for each experimental setup, the averaged momentum flux (${M}$) at the jet exit was estimated and matched for all test cases. This matching was achieved iteratively, by varying the volumetric flow rate (${Q}$) in the system.  Here,  ${M}$ was calculated by first obtaining the time-averaged jet exit velocity from two-dimensional PIV measurements.  The two-dimensional momentum flux, in units of [N/m], was then calculated from
\begin{equation}
	{M}=\int_{-{D}/2}^{{D}/2} {{\rho}_{{j}}  \langle{\boldsymbol{u}({r})}\rangle^{2}  \mathrm{d}{r}}
	\label{eqn.momentum_flux_experiment}
\end{equation}
where the subscript `$j$' refers to the conditions at the nozzle, the angle brackets `$\langle{}$  ${}\rangle$' refer to a time-averaged quantity, and also $\rho$ and $r$ refer to density and radius, respectively. Table \ref{tab.Flow properties_HSlot1} shows the flow properties used in this study, for the horizontal 3D jet configurations, as well as vertical 3D and OP jets used for comparison \citep{Soleimaninia2018IJoHE}; H and V refer to horizontal and vertical orientations, respectively. In all cases, the jets were characterized by the outer-scale Reynolds number, $Re_{\delta}= \langle{\boldsymbol{u}_{j}}\rangle\delta/{\nu}_{\infty}$, where, ${\nu}_{\infty}$ is the ambient fluid kinematic viscosity and $\delta$ is the width of the mean axial velocity profile, evaluated from limits of 5\% of the centreline velocity at $x\simeq0$. 

\subsection{Measurement techniques}
Particle imaging velocimetry (PIV) was used to capture the two-dimensional velocity flow field information. A dual-head Nd: YAG pulsed laser (New Wave's SOLO III  15 HZ) was used to illuminate a two-dimensional cross-section of the flow, which was seeded with Di-Ethyl-Hexyl-Sebacate (DEHS), with a typical diameter of less than 1 $\mu$m, to act as a tracer particle.  The light sheet had an approximate height of 5 cm and thickness of 1 mm.  The field of view of the camera (PIV CCD) was a 40$\times$30 $\textrm{mm}^{2}$ window with an approximate pixel size of 6.5 $\mu$m in physical space. Following the procedure of \cite{Su2003JoFM1}, we estimate this resolution to be comparable to the finest scales of the flow, with respect to the Nyquist criterion. Each pair of images were then processed using LaVision DaVis 8.4 software to calculate the global instantaneous flow velocity field. Following the PIV uncertainty propagation method \citep{Sciacchitano2016MSaT84006}, we estimated conservative uncertainty values of 3\% and 6\% in the time-averaged velocity and Reynolds shear stress profiles, respectively.  

To measure the gas concentration, we applied planar laser-induced fluorescence (PLIF). To simultaneously apply PLIF, the flow was also seeded with acetone vapour at consistent rate of  $\sim10\%$ by volume.  A Pulsed Nd: YAG laser (Spectra-Physics INDI-40-10-HG) was used in order to excite the acetone molecules in a light sheet with an approximate height of 5 cm and a thickness of 350 $\mu$m, which was then recorded with a PLIF CCD camera. The camera field of view for all cases corresponded to a 38$\times$28 $\textrm{mm}^{2}$ window with an approximate pixel size of 6.5 $\mu$m. The images were taken at {a} frequency of 5 Hz and then processed using LaVision DaVis 8.4 software.  Following correcting for errors associated with background noise, fluctuations in cross-sectional laser beam intensity, and laser energy per pulse deviations, one can assume the remaining non-uniformity of the scalar field is due to  signal to noise ratio ($S/N$). The error in the $S/N$ can be estimated from the standard deviation of this ratio in an uniform low signal region of the flow field.   Based on this data, and uncertainty propagation method, we estimated the uncertainty in the time-averaged and variances of concentration field to be conservative values of 4\% and 7\%, respectively. For each experimental case, a total of 500 images were acquired to determine the time-averaged {molar} concentration, {$\langle{X}\rangle$}, and variances, ${X^{\prime^2}}$, fields.  Further details of the experimental procedure can be found in \citep{Soleimaninia2018IJoHE}.

\begin{figure}
	\centering
	a)\includegraphics[scale=0.09,trim={0.1cm 0.1cm 0.1cm 0.1cm},clip]{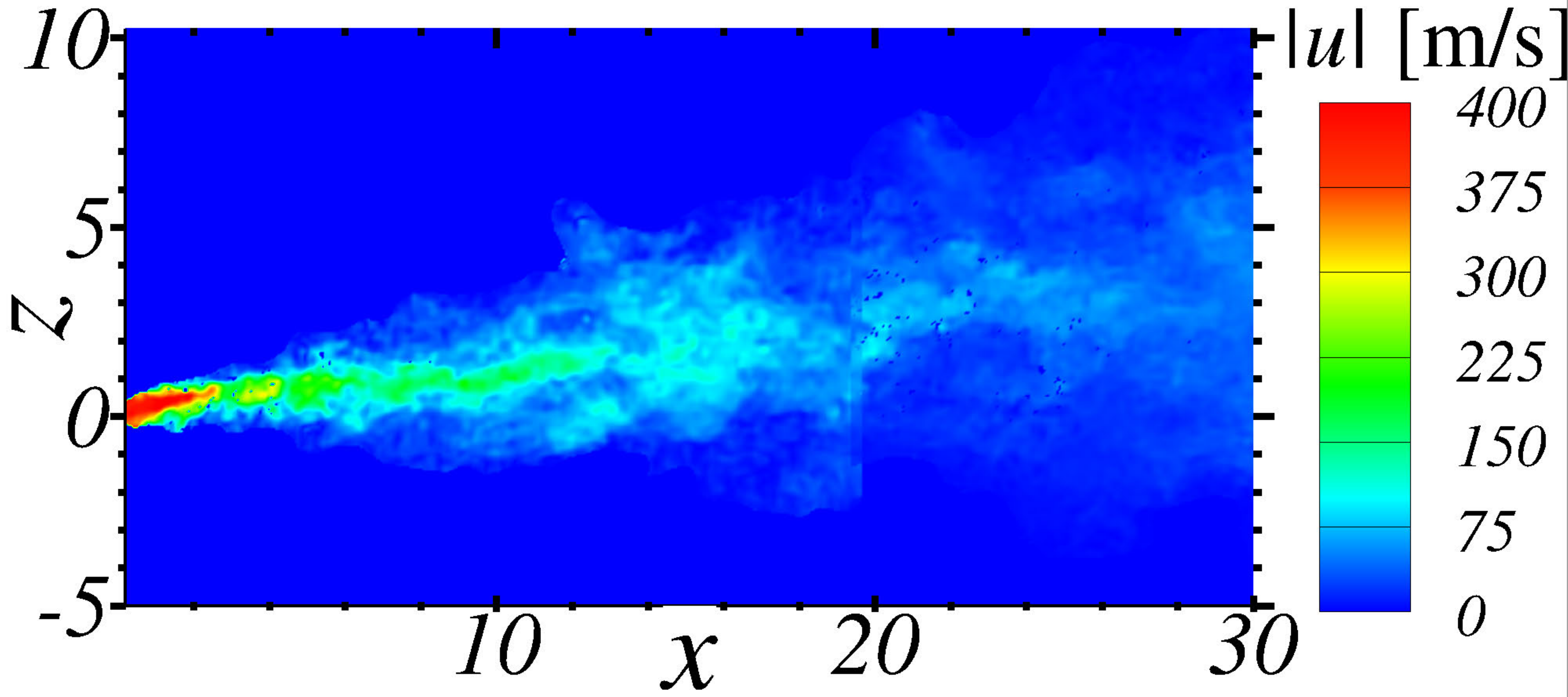}
	b)\includegraphics[scale=0.09,trim={0.1cm 0.1cm 0.1cm 0.1cm},clip]{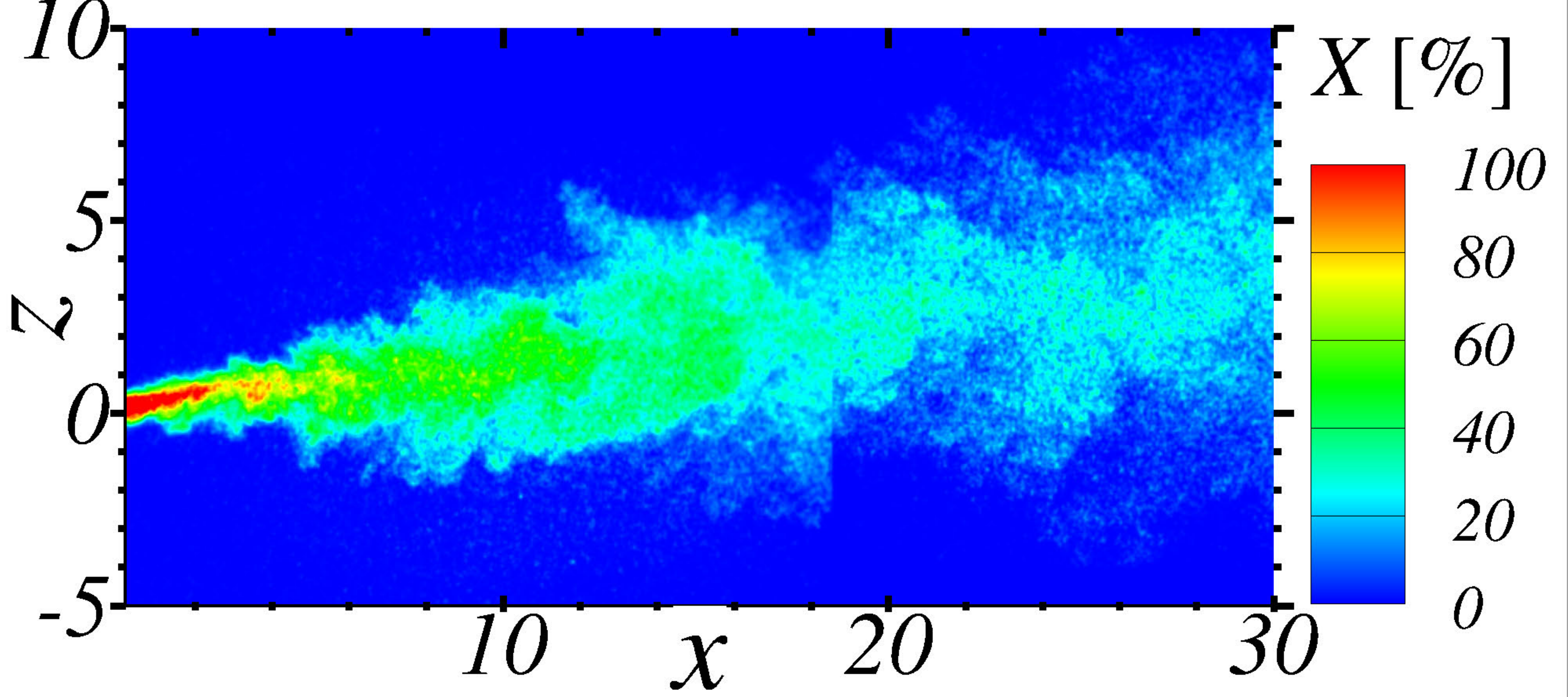}\\
	\caption{Instantaneous a)  velocity and b) molar concentration fields obtained from Helium 3D jet in $x$-$z$ plane from two individual imaging windows stitched together.}
	\label{fig.Instantaneous_plots}		
\end{figure}

Finally, to retain the spatial resolution of the flow field, the full measurement region is covered by two individual imaging windows with at least a 20\% overlap between each window. Figure \ref{fig.Instantaneous_plots} shows examples of the instantaneous velocity and concentration fields, for the helium 3D jet in the $x$-$z$ plane. It should be noted that the flow fields were constructed from two different experiments, where individual imaging windows have been stitched together.

Distances reported here have been normalized such that
\begin{equation}
	x = \frac{\textrm{\bf{X}}}{D}, \;\;\;\; \;\;\;\; y = \frac{\textrm{\bf{Y}}}{D}, \;\;\;\; \;\;\;\; z = \frac{\textrm{\bf{Z}}}{D}, \;\;\;\; \;\;\;\; s = \frac{\textrm{\bf{s}}}{D}, \;\;\;\; \;\;\;\; n = \frac{\textrm{\bf{n}}}{D}
	\label{eqn.Non-Dimensional}
\end{equation}%
where $D$, the diameter of the orifice, is taken as the reference length scale. 

\section{Results}

\subsection{Time-averaged flow fields}
\begin{figure}
	\centering
	\raggedright \underline{\textbf{air:}}\\
	a)\includegraphics[scale=0.09,trim={0.1cm 0.1cm 0.1cm 0.1cm},clip]{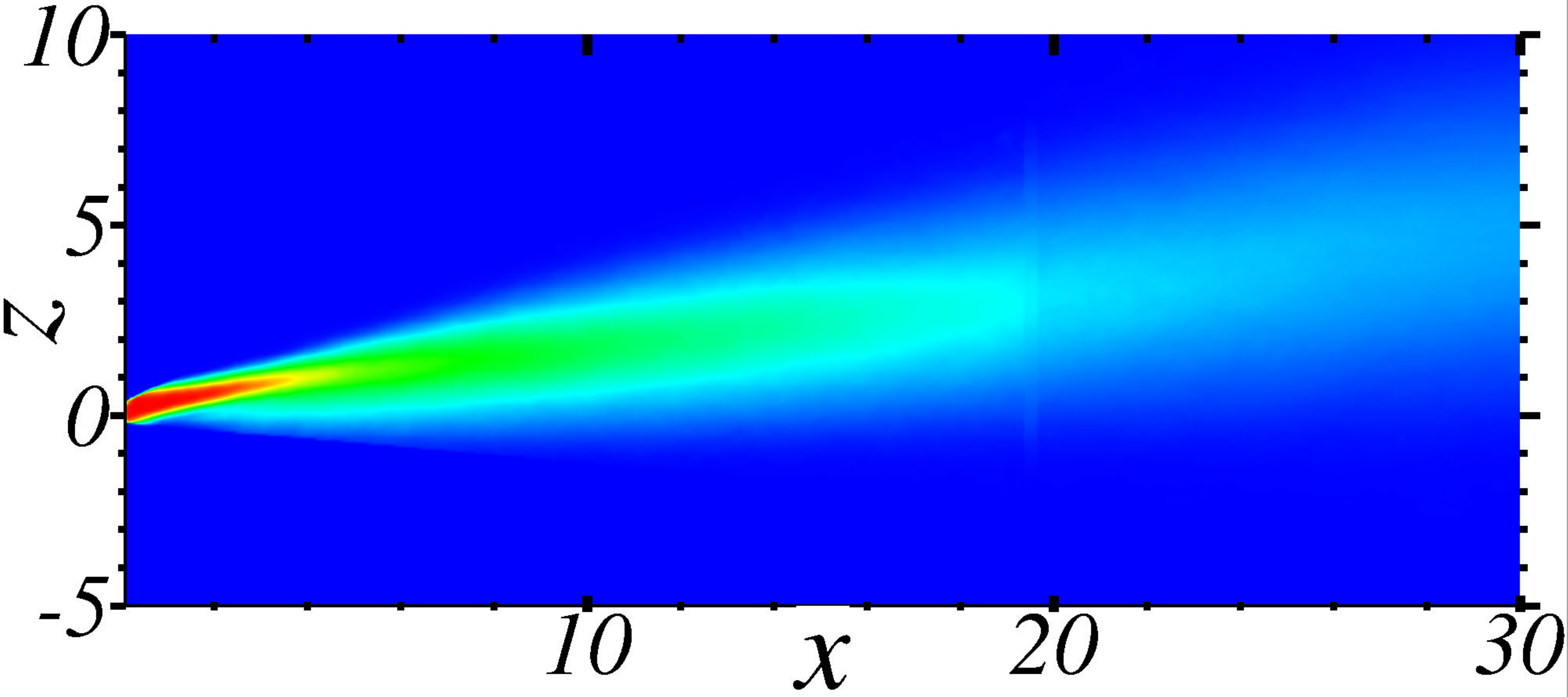}
	b)\includegraphics[scale=0.09,trim={0.1cm 0.1cm 0.1cm 0.1cm},clip]{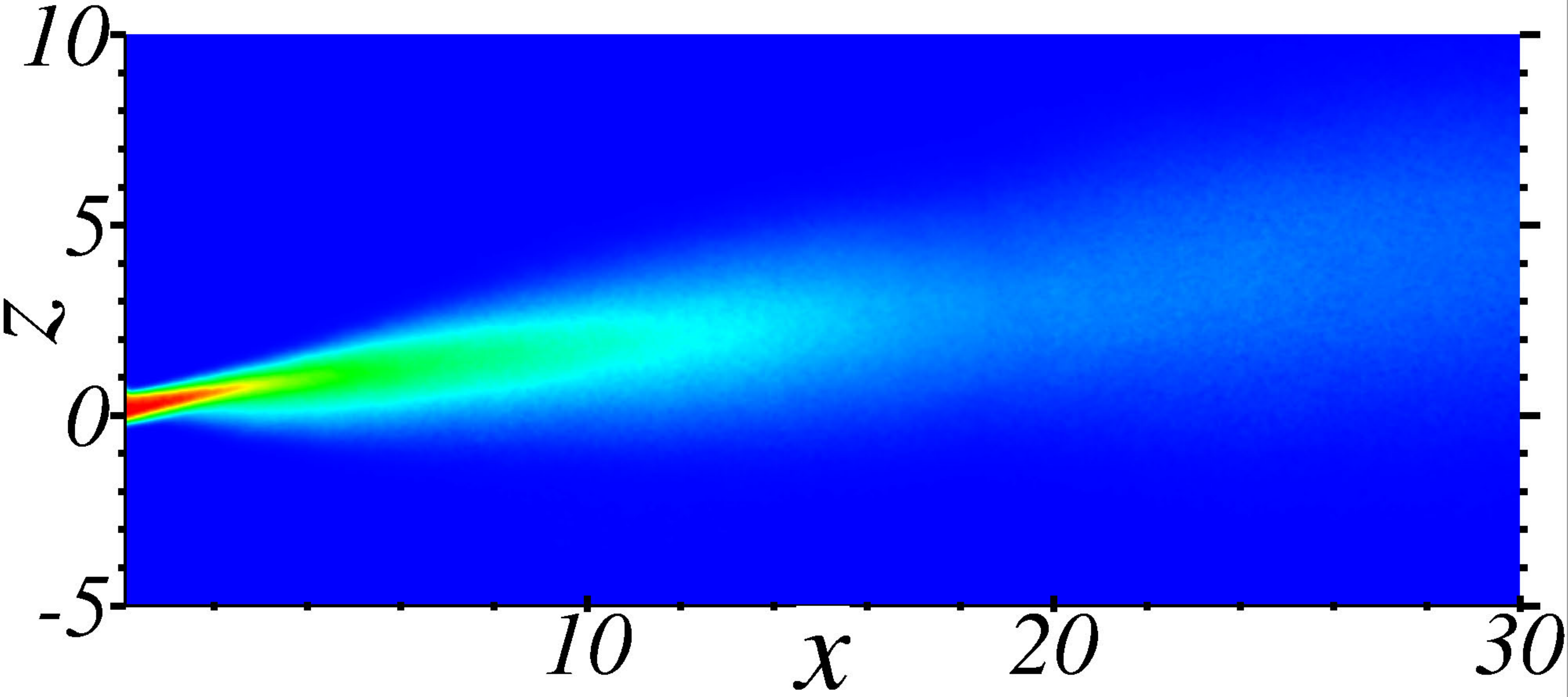}\\
	c)\includegraphics[scale=0.09,trim={0.1cm 0.1cm 0.1cm 0.1cm},clip]{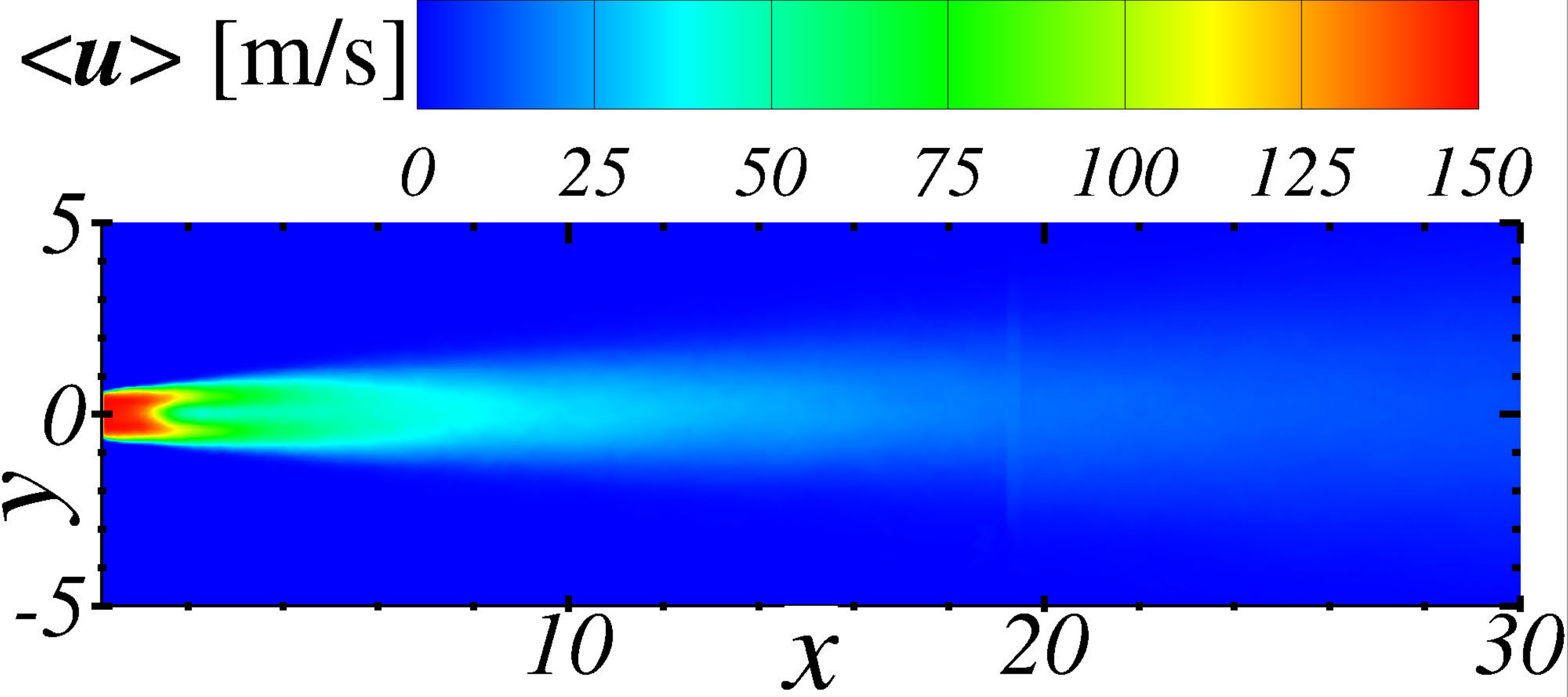}
	d)\includegraphics[scale=0.09,trim={0.1cm 0.1cm 0.1cm 0.1cm},clip]{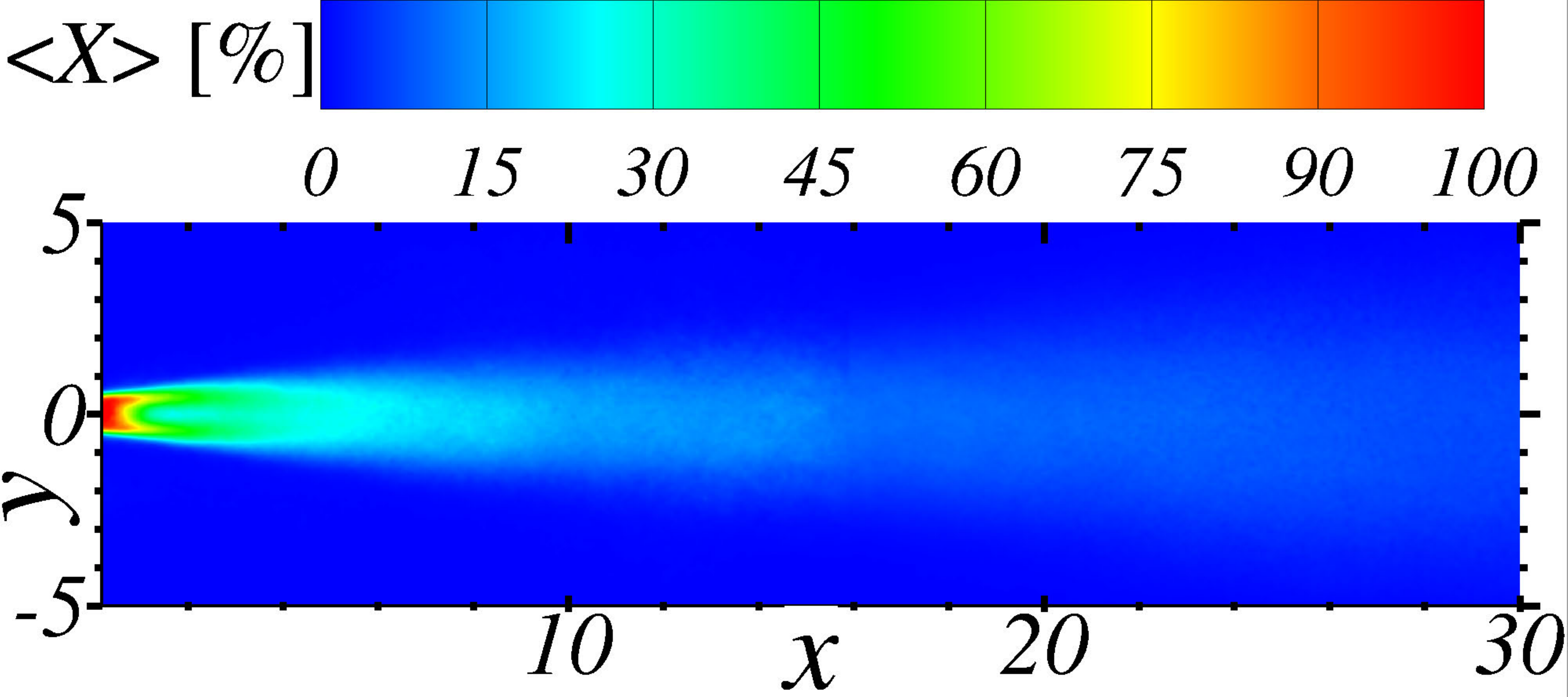}
	\raggedright \underline{\textbf{helium:}}\\
	a)\includegraphics[scale=0.09,trim={0.1cm 0.1cm 0.1cm 0.1cm},clip]{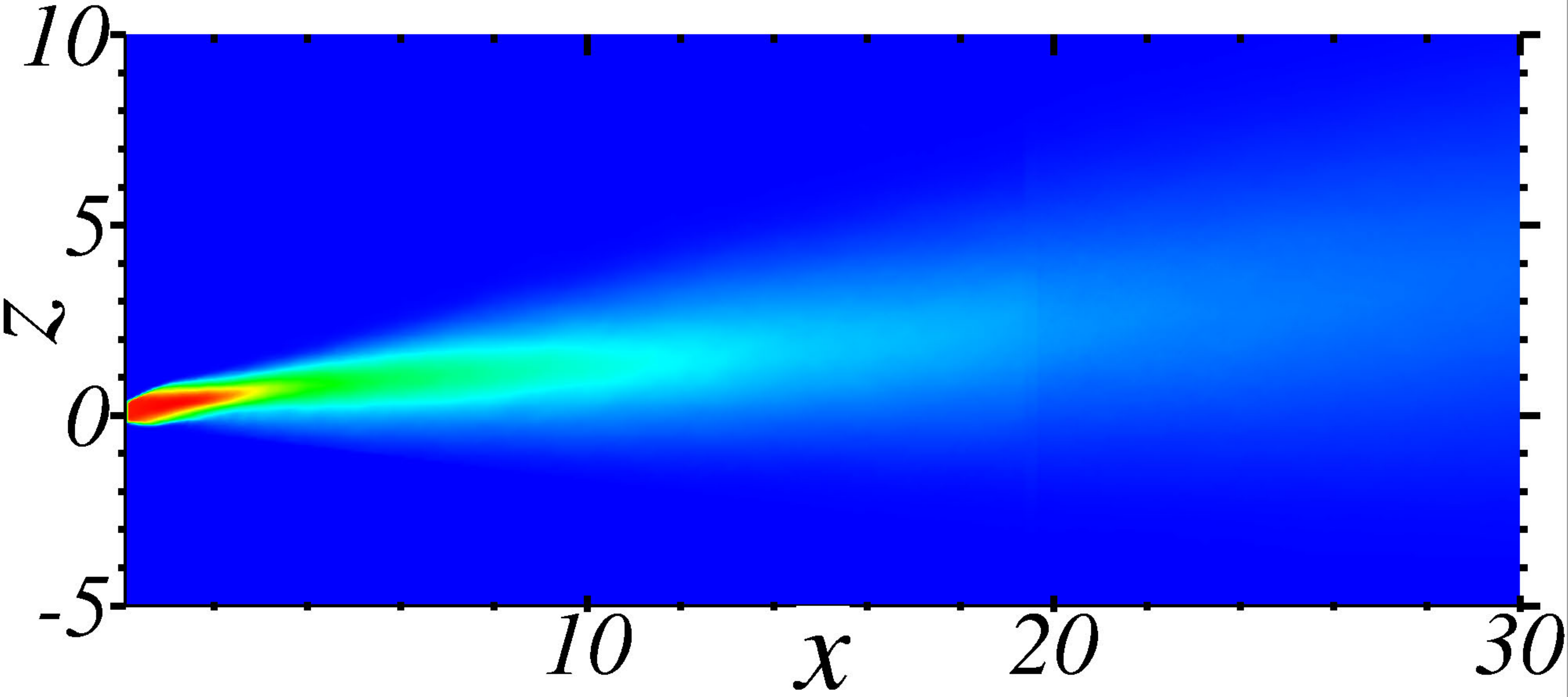}
	b)\includegraphics[scale=0.09,trim={0.1cm 0.1cm 0.1cm 0.1cm},clip]{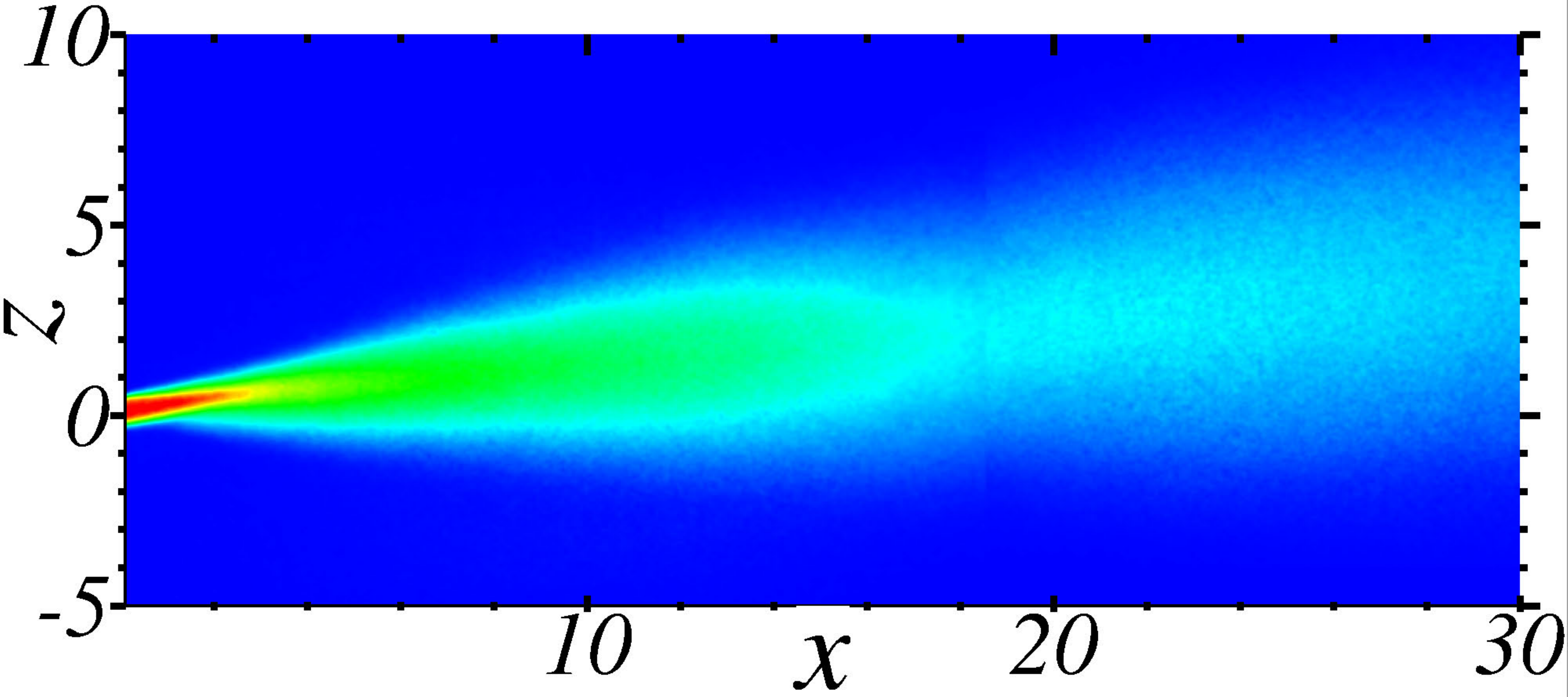}\\
	c)\includegraphics[scale=0.09,trim={0.1cm 0.1cm 0.1cm 0.1cm},clip]{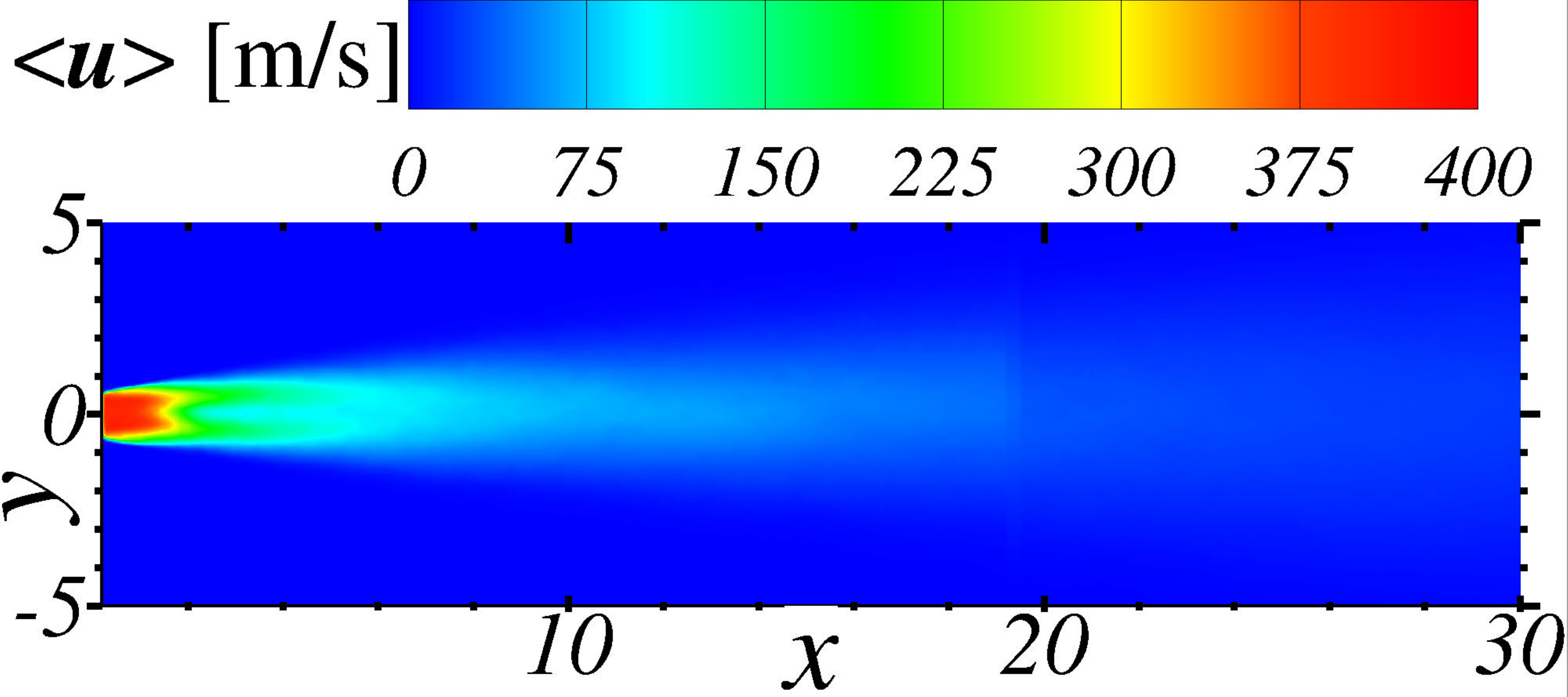}
	d)\includegraphics[scale=0.09,trim={0.1cm 0.1cm 0.1cm 0.1cm},clip]{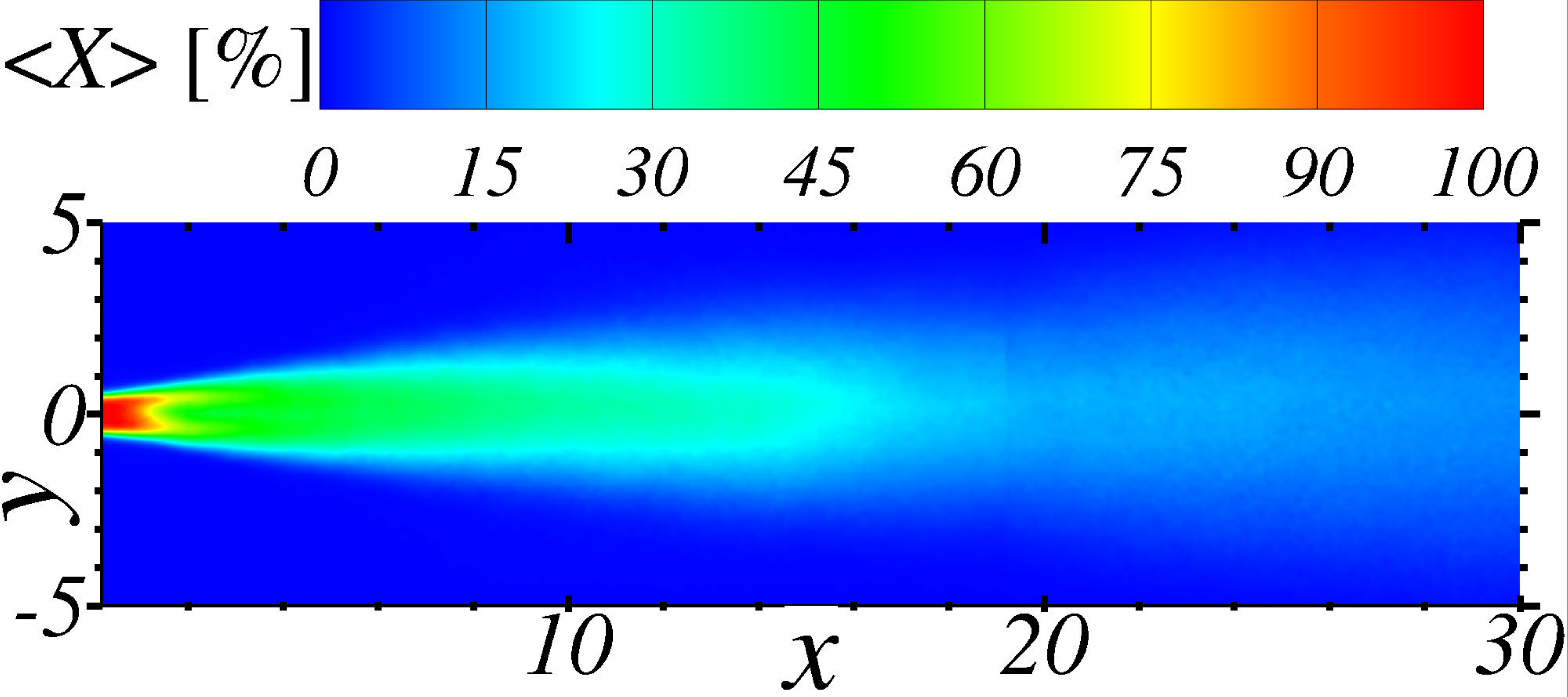}
	\caption{Time-averaged velocity and molar concentration contours from round jet on side of tube (3D jet) for air and helium, obtained from a) velocity contours in $x$-$z$ plane, b) molar concentration contours in $x$-$z$ plane, c) velocity contours in $x$-$y$ plane and d) molar concentration contours in $x$-$y$ plane.}
	\label{fig.Velocity_Concentartion_Contours}
\end{figure}
The time-averaged velocity and molar concentration contours, obtained in both the $x$-$z$ and $x$-$y$ planes for all of the 3D jet experiments conducted here, are shown in Fig.\ \ref{fig.Velocity_Concentartion_Contours}. For both experiments, significantly {larger} jet spreading was observed in the $x$-$z$ planes compared to the $x$-$y$ plane. Clearly, the flow structure was asymmetric in both experiments. The jets were also found to deviate significantly from the horizontal $x$-axis, for both gases in the $x$-$z$ plane. In this plane, near the potential-core region, there was also more jet spreading on the lower side of the jet compared to the top side.   In the $x$-$y$ planes of both gases, there were two high-velocity peaks (saddle-back behaviour), at $y\pm0.5D$, on each side of the $x$-axis, with a much shorter potential-core length ($\simeq2D$) compared to the $x$-$z$ plane. This saddle-back behaviour was previously found to originate from a velocity deficit region which forms inside the orifice due to flow separation as the gas inside the tube encountered the edge of the orifice \citep{Soleimaninia2018IJoHE}. Also, there was a shorter potential-core length observed for helium ($\simeq3D$), compared to air ($\simeq5D$), as observed in the velocity contours of the $x$-$z$ planes.

In general, the concentration profiles were qualitatively similar to the velocity profiles presented in Fig.\ \ref{fig.Velocity_Concentartion_Contours}, with two exceptions. First, the concentration core lengths in both planes were found to be shorter than the velocity potential cores. The concentration core lengths were approximately $\simeq4D$ in the $x$-$z$ plane for both gases, and $\simeq2D$ and  $\simeq1D$, for air and helium, respectively in the $x$-$y$ plane. Also, much higher concentration levels, with higher spreading rates, were observed for helium in the far field compared to air. This observation can be attributed to a low Schmidt number ($Sc<1$), where mass diffusion rates are higher than momentum diffusion rates.

\begin{figure}
	\centering
	\includegraphics[scale=0.35]{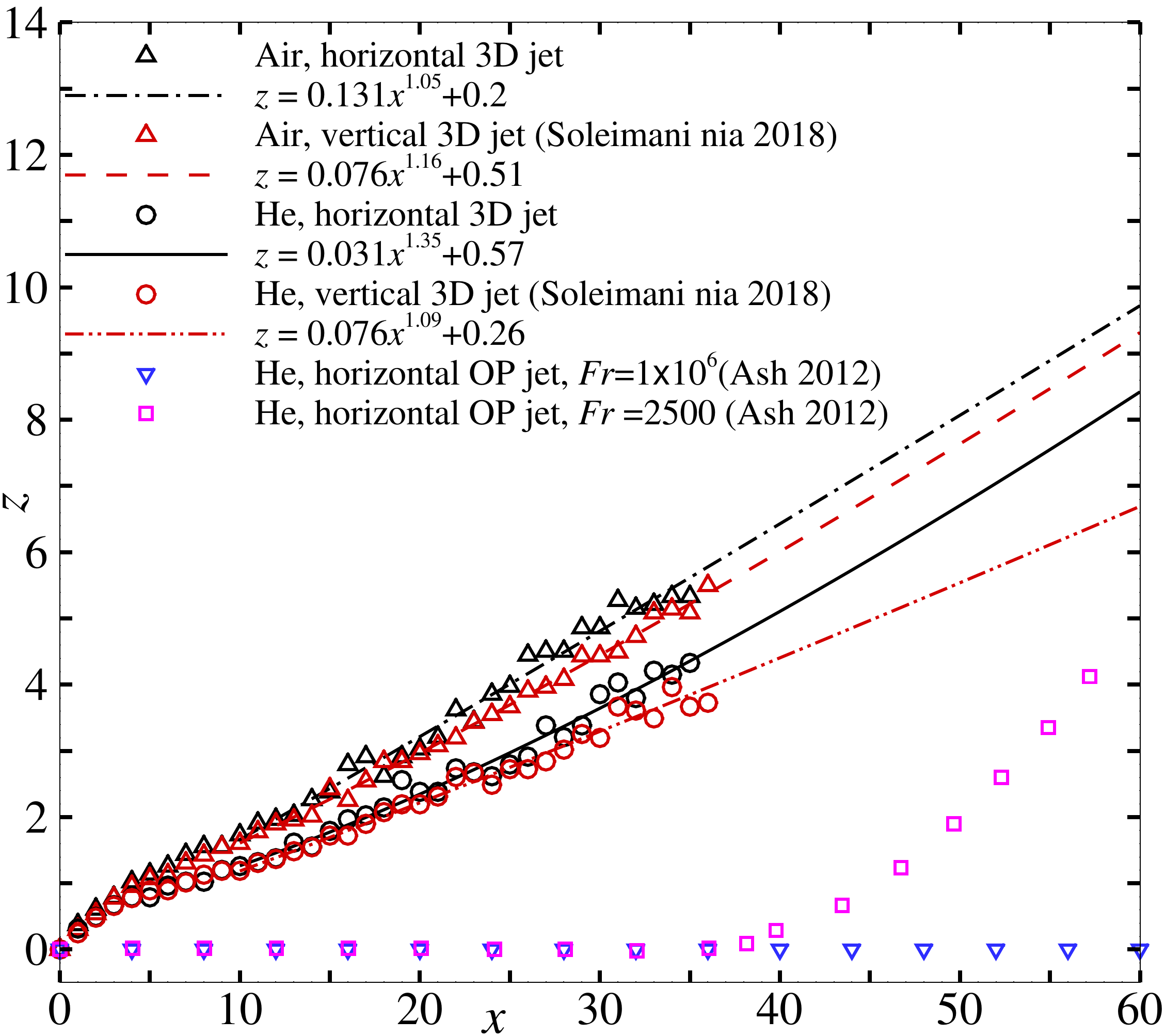}
	\caption{Jet centre-lines taken along the location of maximum velocity {magnitudes ($|\langle{ \boldsymbol{u}}\rangle |_{\textrm{max}}(x)$)} in $x$-$z$ plane from measurements. Also shown for comparison {are} vertical 3D \& OP jets \citep{Soleimaninia2018IJoHE} and horizontal round OP jets experiments \citep{Ash2012}.}
	\label{fig.Jet_Centerline_Trajectory}
\end{figure}
\subsection{The jet centreline trajectory}
In Fig.\ \ref{fig.Jet_Centerline_Trajectory}, the jet centreline trajectories{,} determined in the $x$-$z$ plane, {are} presented for all cases. Here, the trajectories were determined by the maximum velocity magnitude, $|\langle{ \boldsymbol{u}}\rangle |_{\textrm{max}}(x)$, locations. Also shown for comparison are the jet centreline trajectories obtained from previous vertical 3D jet experiments \citep{Soleimaninia2018IJoHE}, and from horizontal sharp-edged orifice flat-plate (OP) helium jet measurements \citep{Ash2012}. In order to determine the effect of buoyancy on the horizontal jets, lines of best fit, using linear regression to power law expressions, were obtained for the far field (beyond $x\geq10D$), and are also shown in Fig.\ \ref{fig.Jet_Centerline_Trajectory}. In general, the jet trajectory for the vertical and horizontal air jets were found to be described by a nearly linear relation (i.e. power law exponent $\sim1$).  The horizontal helium jet, however, was found to have a power law exponent $\sim1.3$.  Upon extrapolating these relations to the far field, beyond the experimental data collected, it {became} clear that buoyancy of the helium jet caused significant deflection from the horizontal axis, despite the high Froude number ($Fr= 1.34\times 10^6$).  It should be noted that for horizontal flat-plate OP helium jets, with a comparable Froude number ($Fr=1\times 10^6$), such buoyancy effects were not observed \citep{Ash2012}.

\subsection{Velocity decay and jet spreading rates}
\begin{figure}
	\centering
	a)\includegraphics[scale=0.27]{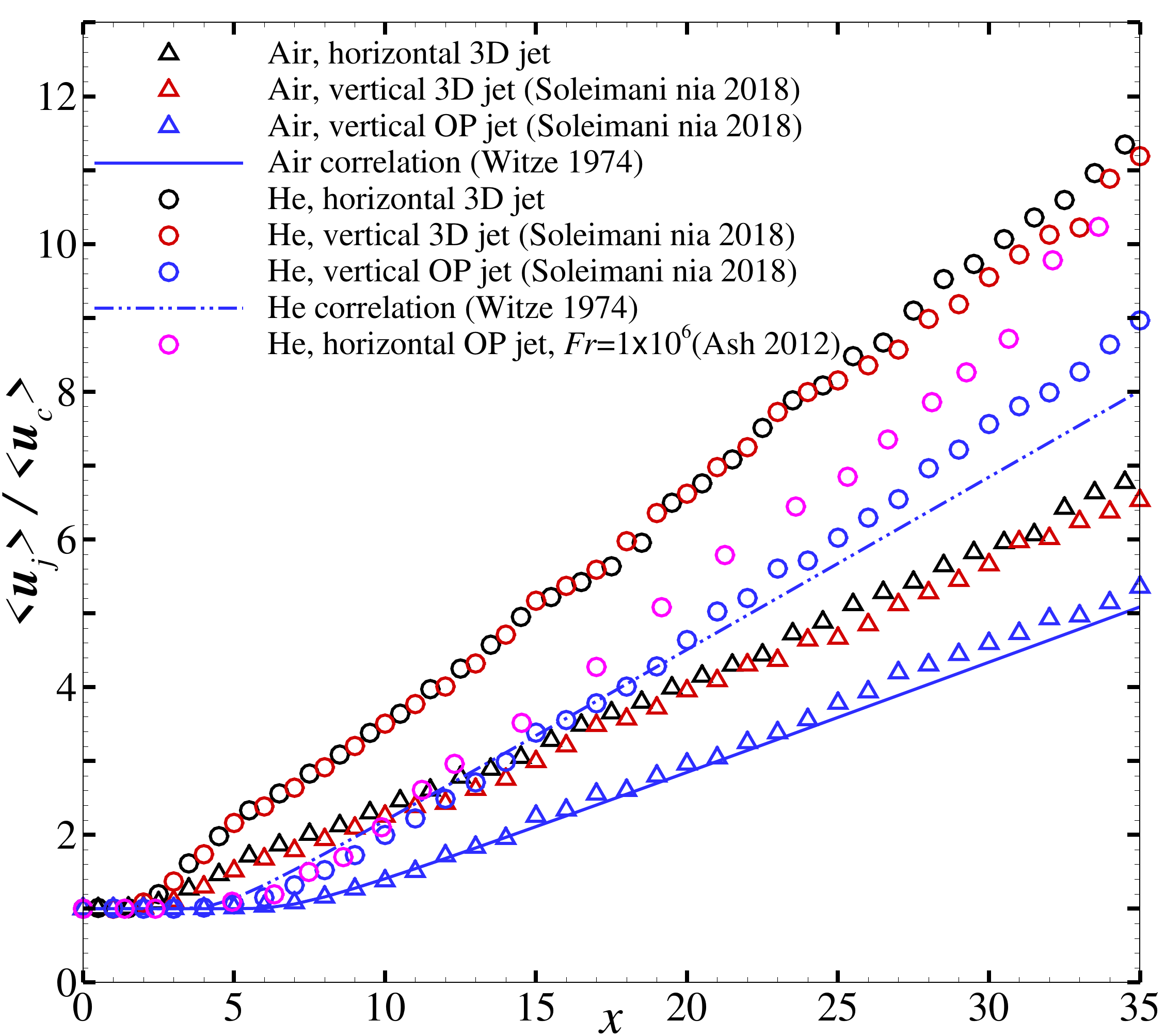} 
	b)\includegraphics[scale=0.27]{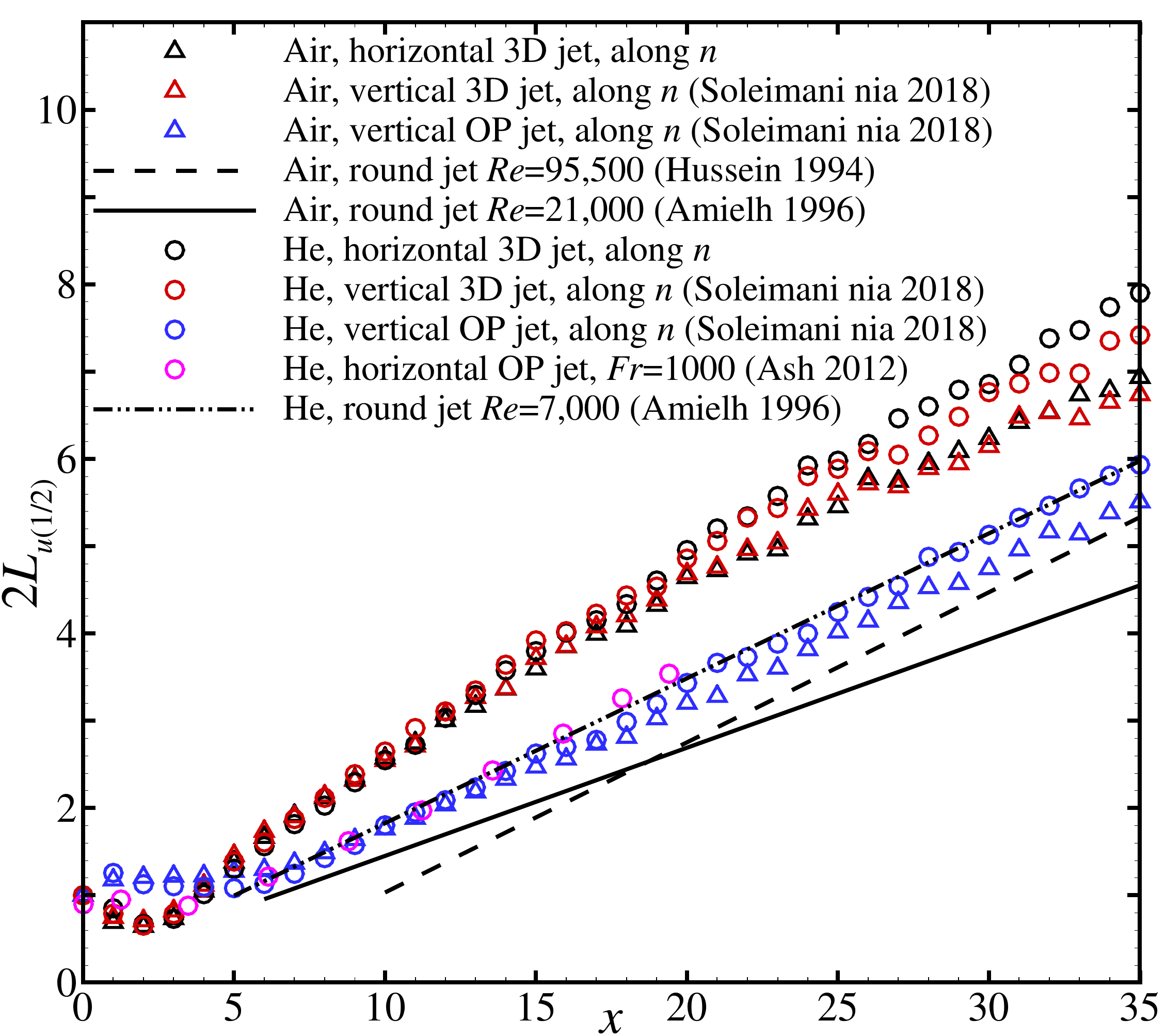} 
	\caption{a) Inverse time-averaged velocity decay and b) jet velocity widths (2$L_{u(1/2)}$) obtained along the $|\langle{ \boldsymbol{u}}\rangle|_{\textrm{max}}(x)$ centrelines, in $x$-$z$ plane from measurements. Note, $n$-coordinate refers to lines which are normal to the centreline, and coplanar with the $x$-$z$ plane (see the coordinate system in Fig.\ref{fig.Experimental_Layout} b). Also shown, for comparison {are} axisymmetric round jet correlations \citep{Witze1974AJ417}, and vertical 3D \& OP jets, horizontal round OP jets and round pipe jet experiments \citep{Soleimaninia2018IJoHE,Ash2012,Hussein1994JoFM31,Amielh1996IJoHaMT2149}.} 
	\label{fig.Velocity_JetDecay_SpreadingRate}
\end{figure}

Fig.\ \ref{fig.Velocity_JetDecay_SpreadingRate}a shows the inverse time-averaged velocity decay ($\langle{ \boldsymbol{u}_j}\rangle/\langle{ \boldsymbol{u}_c}\rangle$) along the jet centrelines ($s$-coordinate illustrated in Fig.\ \ref{fig.Experimental_Layout}b) for all experiments. Here, the subscript `$c$' refers to the conditions at the jet centreline, while the subscript `$j$' refers to the jet exit condition. Also shown, for comparison, are velocity decay correlations \citep{Witze1974AJ417} for compressible subsonic and supersonic axisymmetric round jets, along with velocity decay rates obtained from vertical 3D  and OP jet experiments \citep{Soleimaninia2018IJoHE}, and horizontal OP helium jet measurements \citep{Ash2012}.  Upon comparison to the Witze correlations \citep{Witze1974AJ417}, the air and helium OP jet experiments were found to reproduce well the expected velocity decay rate, with helium {jet} decaying faster than the air jet. On the other hand, the decay rates observed in the experimental 3D jets were much faster compared to the axisymmetric jets. In general, upon comparison between horizontal and vertical 3D jets, buoyancy was not found to significantly affect the velocity decay rates.\\

In the $x$-$z$ plane, Figure \ref{fig.Velocity_JetDecay_SpreadingRate}b presents the jet velocity widths (2$L_{u(1/2)}$), that have been obtained by determining the locations where $|\langle{ \boldsymbol{u}}\rangle|=0.5|\langle{ \boldsymbol{u}}\rangle|_{\textrm{max}}(x)$ along lines which were orthogonal to the jet-centrelines.  These orthogonal lines have been indicated previously as coordinate `$n$' in Fig.\ \ref{fig.Experimental_Layout}b. For the 3D jets, in all cases, a slight contraction in the jet widths has been observed from $1D<x<4D$.  Beyond this point, the jet spreading rates, along $n$, were observed to be much greater compared to the axisymmetic jets for all cases.  Moreover, the air and helium jet spreading, from the 3D experiments, was found to be comparable for both gases.  However, In the far field (beyond $x\geq13D$), the helium 3D jets exhibited higher spreading rates, compared to air. This trend was slightly more clear upon comparison to the horizontal 3D cases between helium and air. In general, the OP jets were found to have nearly constant jet widths in the potential core region, up until {$x\sim5D$}.  From this point on, the OP jet widths were found to be much smaller compared to 3D jets, with the expected linear increase in jet spreading rates of previous axisymmetric round jet experiments for a wide range of Reynolds numbers \citep{Hussein1994JoFM31,Amielh1996IJoHaMT2149}. 

\subsection{Scalar concentration decay and jet spreading rates}

\begin{figure}
	\centering
	a)\includegraphics[scale=0.27]{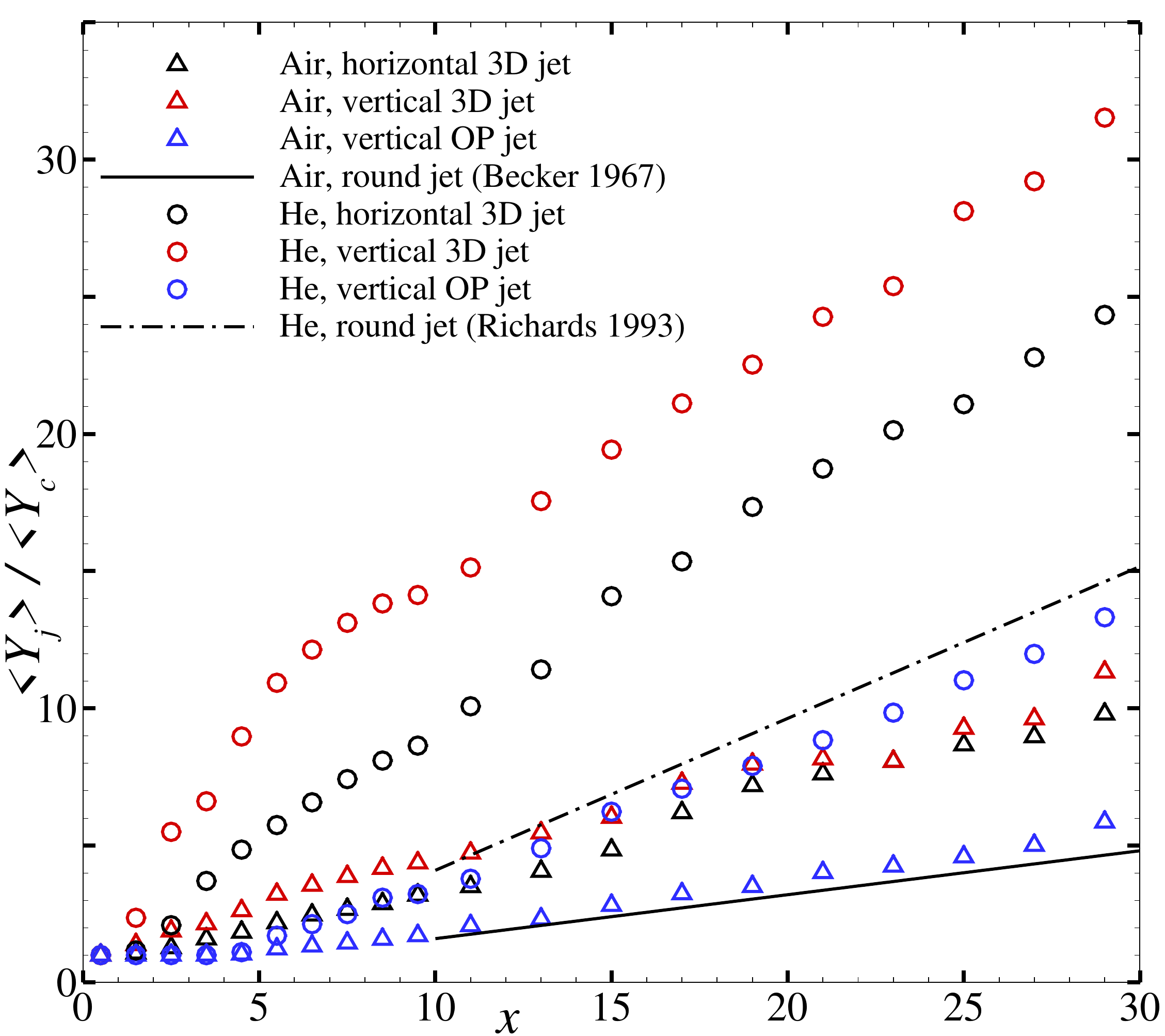} 
	b)\includegraphics[scale=0.27]{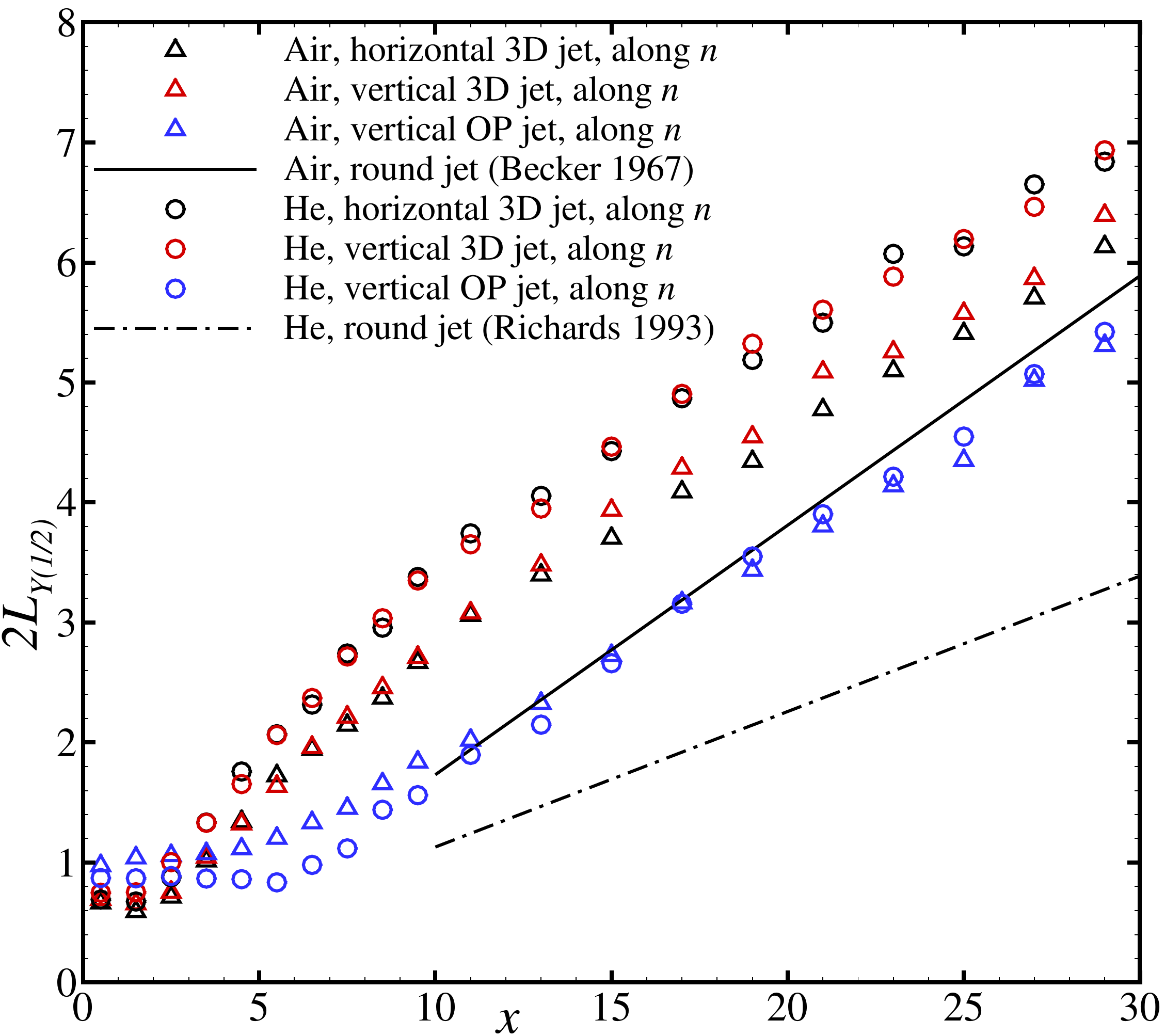}	
	\caption{a) Inverse time-averaged jet gas mass fraction decay and b) mass fraction jet widths (2$L_{Y(1/2)}$) obtained along the $\langle{Y}\rangle_{\textrm{max}}(x)$ centrelines, in $x$-$z$ plane from measurements. Also shown, for comparison {are} vertical 3D \& OP jets, and round pipe jet experiments \citep{Soleimaninia2018IJoHE,Becker1967JoFM285,Richards1993JoFM417}.} 
	\label{fig.Concentration_JetDecay_SpreadingRate}
\end{figure}

Figure\ \ref{fig.Concentration_JetDecay_SpreadingRate}a shows centreline evolution of the inverse time-averaged jet gas mass fraction, $\langle{Y_j}\rangle/\langle{Y_c}\rangle$, for both air and helium measurements. Here, the jet gas mass fractions were determined from the measured mole fractions through
\begin{equation}
	Y=\frac{X W_j}{\overline{W}}
	\label{eqn.mass_fraction}
\end{equation}
where $X$ and $W_j$ refer to the mole fraction and molecular weight of the jet gas, respectively, and $\overline{W}$ refers to the mean molecular weight of the local jet gas-ambient air mixture given by
\begin{equation}
	\overline{W}= X W_j + (1-X) W_{\textrm{air}}
	\label{eqn.mean_molecular_weight}
\end{equation} 

Also shown, for comparison, are the centreline mass fraction decay rates for axisymmetry round air \citep{Becker1967JoFM285} and helium jets \citep{Richards1993JoFM417}, along with the mass fraction decay rates obtained from vertical 3D  and OP jet experiments \citep{Soleimaninia2018IJoHE}. In general, the air and helium vertical OP jet experiment mass fraction decay rates compared well to previous axisymmetry round pipe jet experiments \citep{Becker1967JoFM285,Richards1993JoFM417}, where helium jets were always observed to decay faster than air jets. It is noted, however, that the slight differences observed in decay rates for the axisymmetric helium jets (OP and pipe jets) are likely due to differences in the Reynolds numbers between experiments ($Re=4,000$ for the round pipe jets compared to $Re=44,200$ for the OP helium jet) \citep{Pitts1991EiF135}.  Differences in the geometry of the jet outflow condition may have also been a factor. Also, the centreline mass fraction decay rates observed in the experimental 3D jets were much faster compared to the axisymmetric jets. Moreover, upon comparison of the 3D helium jets, the vertical orientation was found to have a faster mass fraction decay rate compared to the horizontal case.  Such differences in behaviour was not observed for the 3D air jets, suggesting that buoyancy plays a significant role on the mass fraction decay rates. Also, upon comparison to the velocity decay rates in Fig.\ \ref{fig.Velocity_JetDecay_SpreadingRate}a, we note that the jet centerline mass fraction decays faster than the velocity for helium, owing to the low Schmidt number ($Sc<1$).

As was done for the velocity field, the jet widths based on the jet gas mass fraction (2$L_{Y(1/2)}$) have been obtained for each experiment, and presented in Fig.\ \ref{fig.Concentration_JetDecay_SpreadingRate}b. This was achieved by determining the locations where $\langle{Y}\rangle=0.5\langle{Y}\rangle_{\textrm{max}}(x)$ along orthogonal lines to the jet centreline, in the $x$-$z$ planes. For all 3D jet cases, a slight contraction in the jet mass fraction widths was observed from {$x<4D$}, as was previously observed for the jet widths based on velocity. Beyond this point, the jet scalar growth rates, along $n$, were observed to be much greater compared to the axisymmetic jets for all cases. The helium jet also exhibited a faster spreading rate compared to air, in both horizontal and vertical cases. The air and helium OP jets were found to have nearly constant mass fraction widths in the potential core region, up until {$x\sim5D$}. After the potential core region, the jet scalar width was found to be much smaller compared to 3D jets, and increase linearly, consistent with the jet mass fraction spreading rates of previous axisymmetric round jet experiments \citep{Becker1967JoFM285,Richards1993JoFM417}.

\subsection{Jet centreline statistics}

\begin{figure}
	\centering
	a)\includegraphics[scale=0.27]{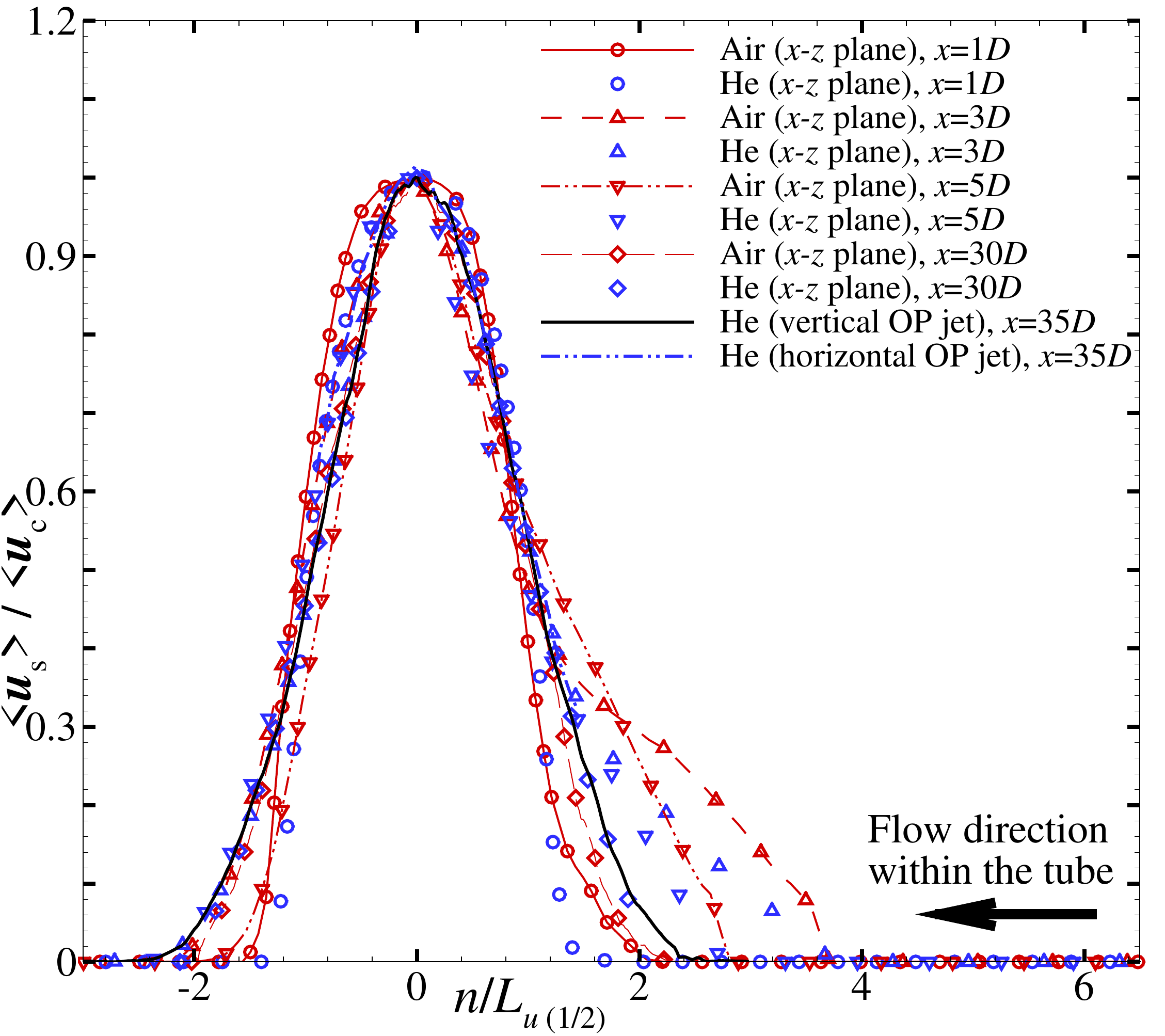} 
	b)\includegraphics[scale=0.27]{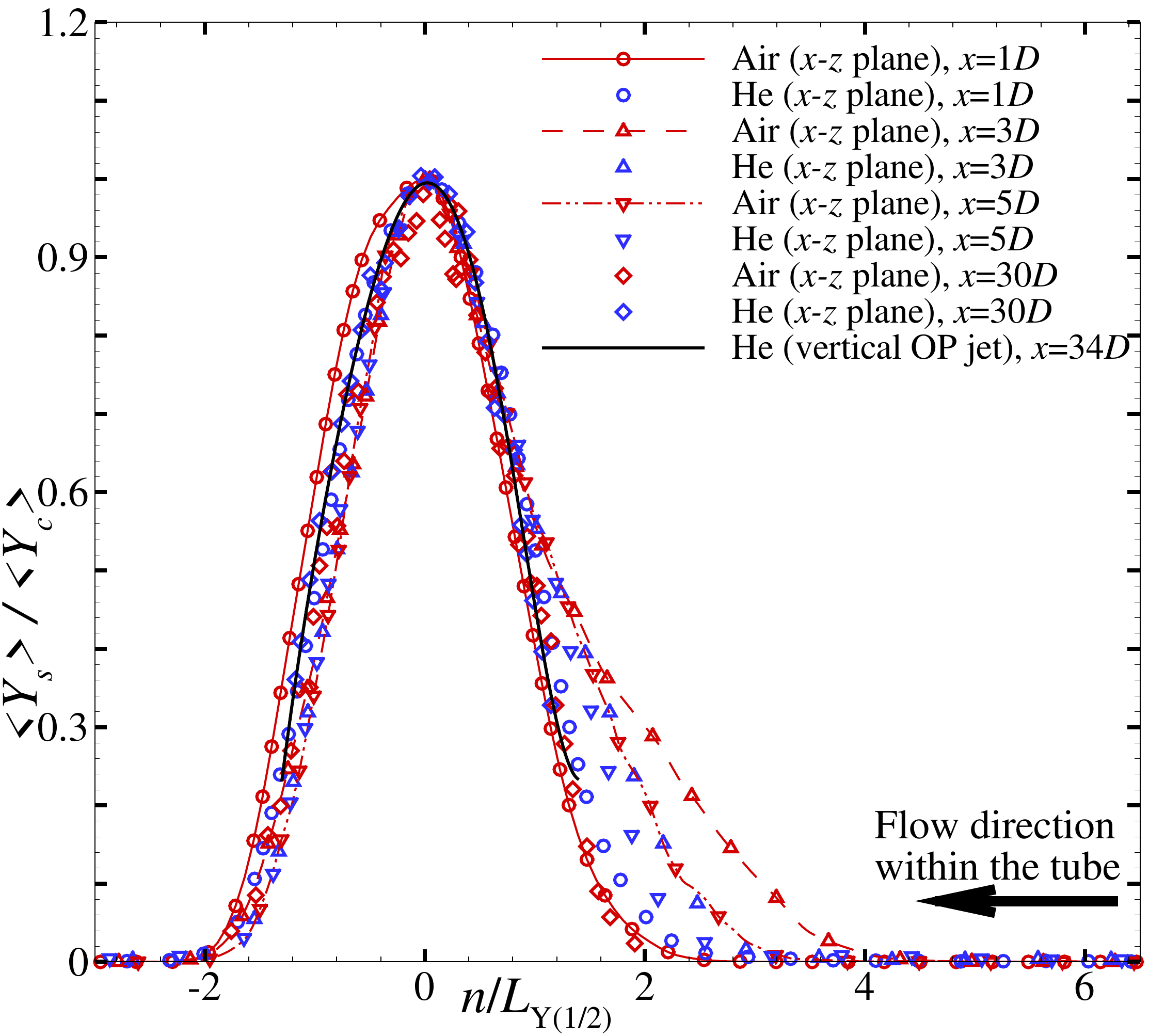}\\
	\caption{a) Normalized time-averaged velocity, and b) concentration profiles along jet centrelines in $x$-$z$ plane, taken at various heights for both air and helium. Time-averaged velocity and concentration profiles are also compared to experimental axisymmetry horizontal ($Fr=1\times 10^6$)\citep{Ash2012} and vertical round OP jets \citep{Soleimaninia2018IJoHE}.}
	\label{fig.Vs_Cs}
\end{figure}
In the $x$-$z$ plane, the normalized time-averaged $s$-velocity components and jet gas mass fraction profiles, for all 3D and OP jet experiments, are shown in Fig.\ \ref{fig.Vs_Cs} along the $n$-coordinate (see Fig.\ \ref{fig.Experimental_Layout}b) for several downstream locations along the jet centreline ($s$-curve in Fig.\ \ref{fig.Experimental_Layout}b). It should be noted that the $s$-component velocities, presented here, were normalized by the local centreline velocity magnitudes,  $\langle{ \boldsymbol{u}}\rangle_{c}(s)$.  The time-averaged jet gas mass fractions ($\langle{Y}\rangle$) have been normalized by the local centreline jet gas mass fraction, $\langle{Y_c}\rangle(s)$.  Also in the figure, the $n$-coordinates which are normal to the centreline $s$-curve, were normalized by the jet velocity half widths ($L_{u(1/2)}$) and the jet gas mass fraction half widths ($L_{Y(1/2)}$), respectively. In general, all 3D jet cases developed into a self-similar Gaussian-like distribution of velocity within the range $|n/L_{1/2}| < 1$ for {$x\le5D$}. However, notable deviations from the self-similar solution were observed near the tail ends of the curves in the $x$-$z$ plane, beyond this range, especially in the opposite stream wise direction of the flow within the tube (+$n$ direction).  For vertical 3D jets, this was previously found to be due enhanced mixing associated with the original flow orientation relative to the orifice, and also curvature of the tube \citep{Soleimaninia2018IJoHE}. Both air and helium experiments were found to exhibit significantly more velocity and jet gas mass fraction spreading to the lower side of the jet centre (in the $+n$ direction), with more spreading observed in this region for helium compared to air.  Beyond $x>5D$, in the far field, the experimental 3D air and helium jets developed into the self-similar Gaussian distribution obtained from the OP jets for the full range of $n$.

\begin{figure}
	\centering
	a)\includegraphics[scale=0.27]{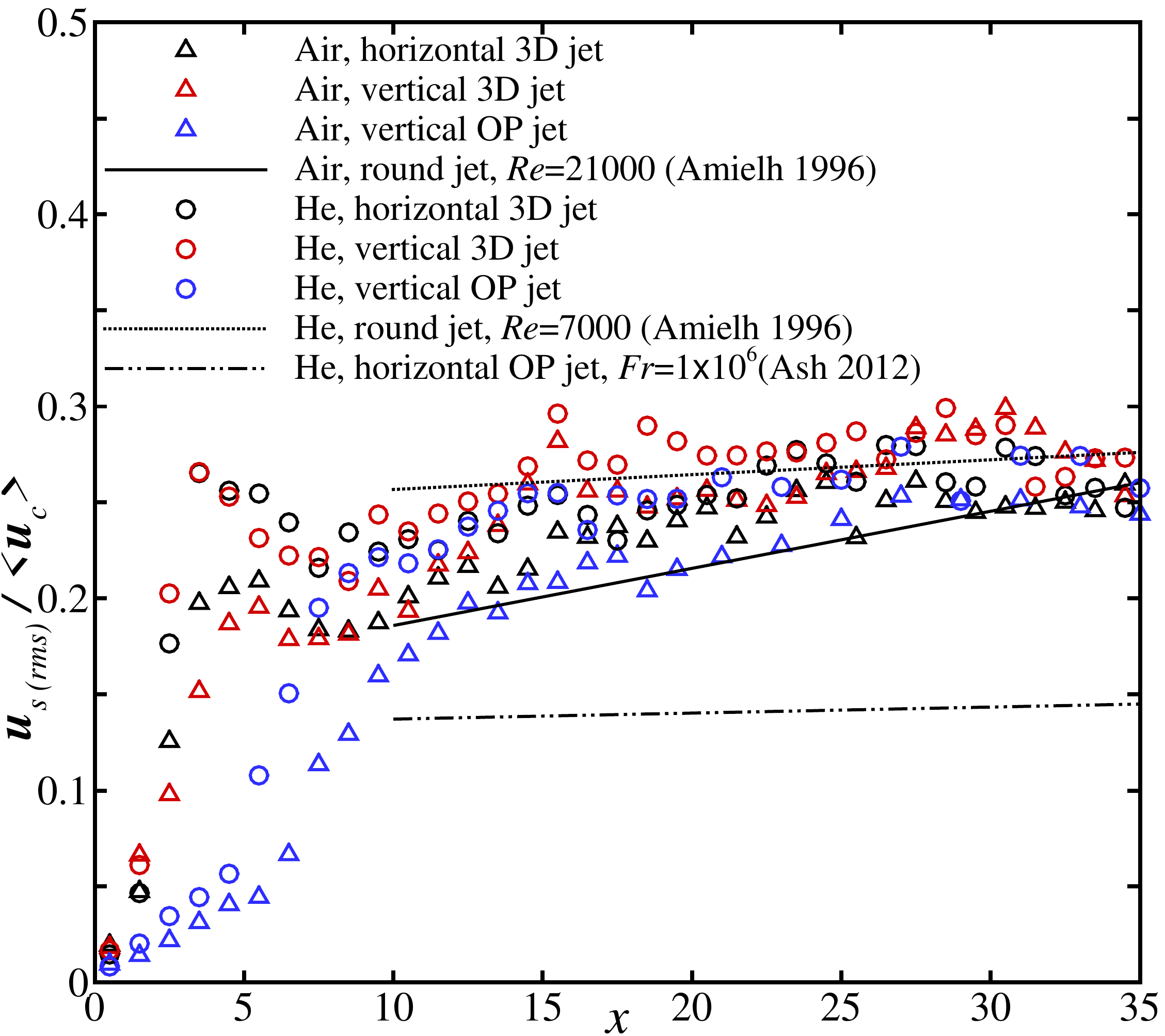} 
	b)\includegraphics[scale=0.27]{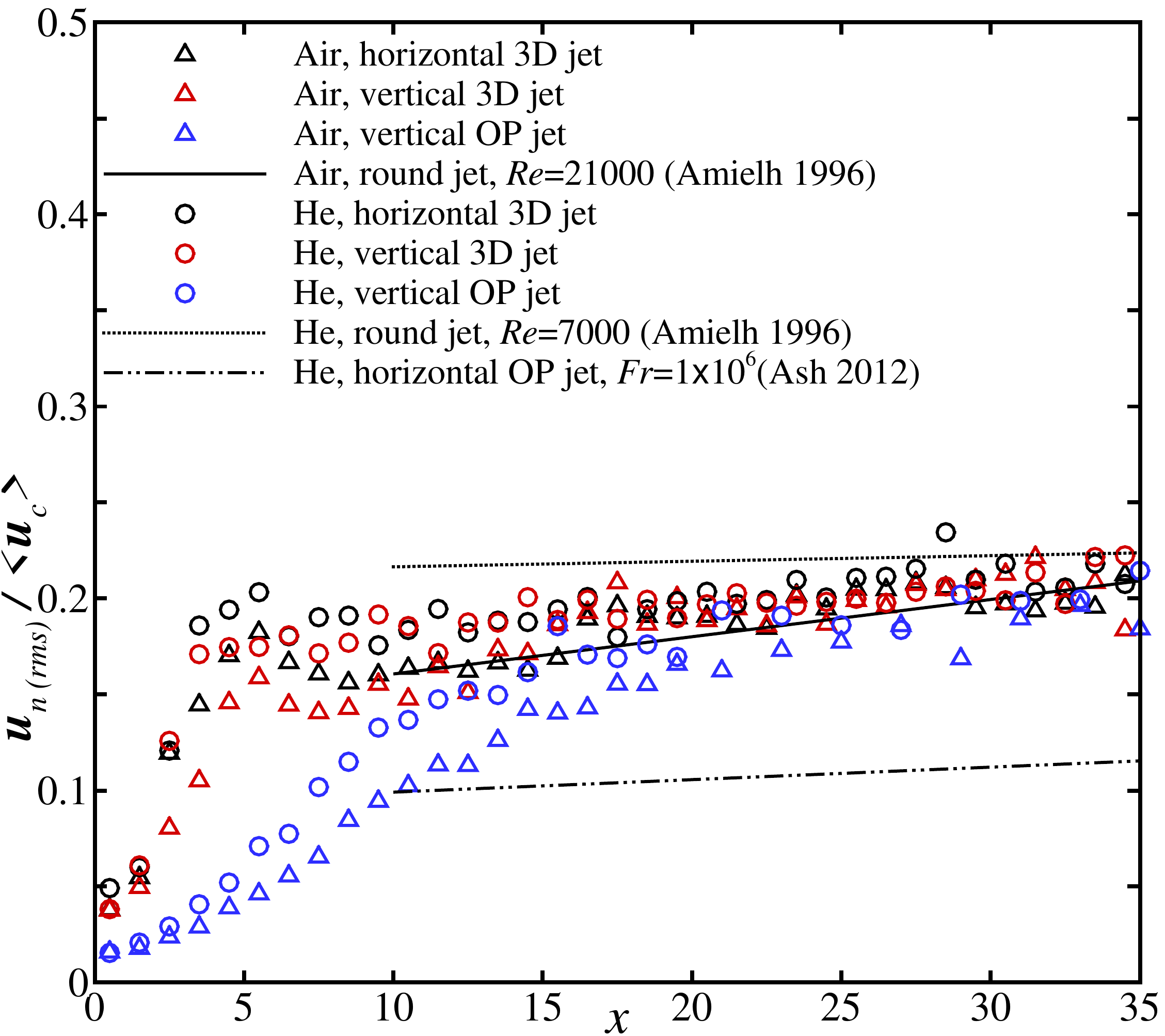}\\
	\caption{Axial development of turbulence intensities along jet centrelines, a) tangential turbulence intensity component ($\boldsymbol{u}_{s(rms)}/ \langle{\boldsymbol{u}_{c}}\rangle$) and b) orthogonal turbulence intensity component ($\boldsymbol{u}_{n(rms)}/ \langle{\boldsymbol{u}_{c}}\rangle$) for experiments. Also shown, for comparison {are} vertical 3D \& OP jets, horizontal OP jet, and round pipe jet experiments \citep{Soleimaninia2018IJoHE, Ash2012, Amielh1996IJoHaMT2149}.}
	\label{fig.Velocity_fluctuation_rms}
\end{figure} 

Fig.\ \ref{fig.Velocity_fluctuation_rms} shows the normalized axial evolution of the r.m.s. velocity fluctuation components in the $s$ and $n$ directions, tangential and orthogonal to jet centreline, where  $\boldsymbol{u}_{(rms)}=\langle{\boldsymbol{u}^{\prime 2}}\rangle^{1/2}$. It should be noted that the prime ($^\prime$) represents the instantaneous fluctuating quantity ($\boldsymbol{u}^{\prime}=\boldsymbol{u}-\langle{\boldsymbol{u}}\rangle$). For the 3D vertical and horizontal helium jets, the tangential turbulence intensity reached an asymptotic value of $\sim26\%$ at $x=3D$, whereas such a value was not observed until $x=20D$ and $x=15D$ for the 3D horizontal and vertical air jets, respectively. This trend was also observed in pipe jet measurements \citep{Amielh1996IJoHaMT2149} and also the current vertical OP jets, where helium reached the asymptotic value closer to orifice, at $x=15D$, compared to air at $x=32D$. However, it appears that this asymptotic value of $\sim26\%$ would be reached in the far field ($x>33D$) for all jets, except the horizontal helium OP jet measurements \citep{Ash2012}. It should be noted that lower turbulent intensities of the horizontal helium OP jet, observed in both tangential and orthogonal components, are likely due to higher initial turbulent intensities reported for the horizontal helium OP jet (not shown)\citep{Mi2007EiF625}. Also, lower spatial resolution of the PIV measurement compared to the current experiments (almost 3 times less), may have been a factor. The same remark is valid for the orthogonal turbulence intensity, as the 3D helium jets reached the asymptotic value of $\sim19-22\%$ more closer to orifice at $x=5D$, compared to air at $x=15D$. Also, the OP vertical helium jet reached this peak turbulence intensity at $x=15D$, whereas such turbulence intensity was not recovered until $x=30D$ for air. In general, the intensity of tangential velocity fluctuations was higher than the orthogonal components, as observed in previous studies \citep{Panchapakesan1993JoFM225, Amielh1996IJoHaMT2149}.

\begin{figure}
	\centering
	\includegraphics[scale=0.35]{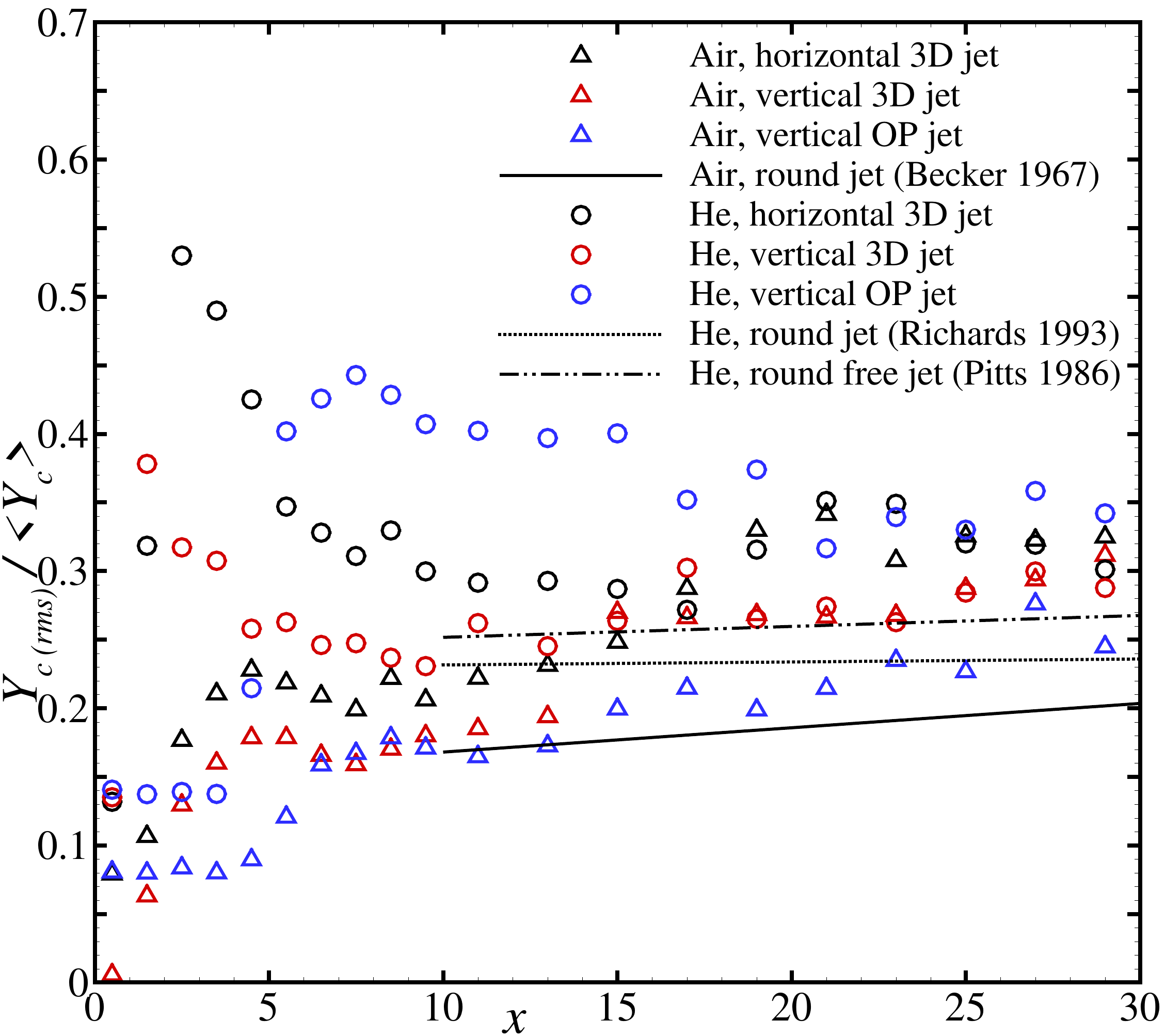} 
	\caption{Normalized axial evolution of mass fraction fluctuation intensities along jet centrelines, ${Y}_{c(rms)}/ \langle{{Y}_{c}}\rangle$, for experiments. Also shown, for comparison {are} vertical 3D \& OP jets, and round pipe jet experiments \citep{Soleimaninia2018IJoHE,Becker1967JoFM285,Pitts1991EiF125,Richards1993JoFM417}.}
	\label{fig.Concenration_fluctuation_rms}
\end{figure} 

Fig.\ \ref{fig.Concenration_fluctuation_rms} shows the normalized axial evolution of the r.m.s.\ jet gas mass fraction fluctuations (unmixedness), ${Y}_{c(rms)}/ \langle{{Y}_{c}}\rangle$, along the jet centreline, for all experiments. In the vertical 3D jets, helium reached the asymptotic value of $\sim26\%$ at $x=5D$, whereas such a value was not recovered until $x=14D$ for air.  This value is in good agreement of the asymptotic value previously reported in helium free jets \citep{Pitts1986}. Also, in horizontal 3D jets, helium reached an umixedness value of $\sim33\%$ at $x=5D$, but then decreased to the asymptotic value of the vertical jets ($\sim26\%$) at $x=17D$, and then again increased to the asymptotic value of $\sim33\%$ for the rest of the measurement domain. For the horizontal 3D air jet, the unmixedness reached a value of $\sim20-24\%$ at $5D<x<15D$, which is in good agreement with the values reported in literature for the far field ($\sim21-24\%$) \citep{Panchapakesan1993JoFM225, Chen1980, Richards1993JoFM417}. Then, the unmixedness recovered the asymptotic value of $\sim33\%$ at $x=19D$ for the rest of observation domain. For the vertical OP air jet, the observed profile followed closely the values of those reported for smooth contraction (SC) air jets \citep{Becker1967JoFM285} in the near field ($5D<x<15D$), and then increased to the asymptotic value of $\sim23\%$ for the rest of the domain. On the other hand, for the vertical OP helium jet, the unmixedness reached a peak value of 0.43 at $x=7D$, then slowly decreased to the value of $\sim33\%$ in the far field, from $x>20D$. it should be noted that, even though the unmixedness values were not consistent between the helium and air OP jets in the measurement domain, extrapolation of the data (not shown) revealed that the far field unmixedness would converged to the same value at about $x>50D$.
\begin{figure}
	\centering
	\raggedright \underline{\textbf{air:}}\\
	\centering
	a)\includegraphics[scale=0.27]{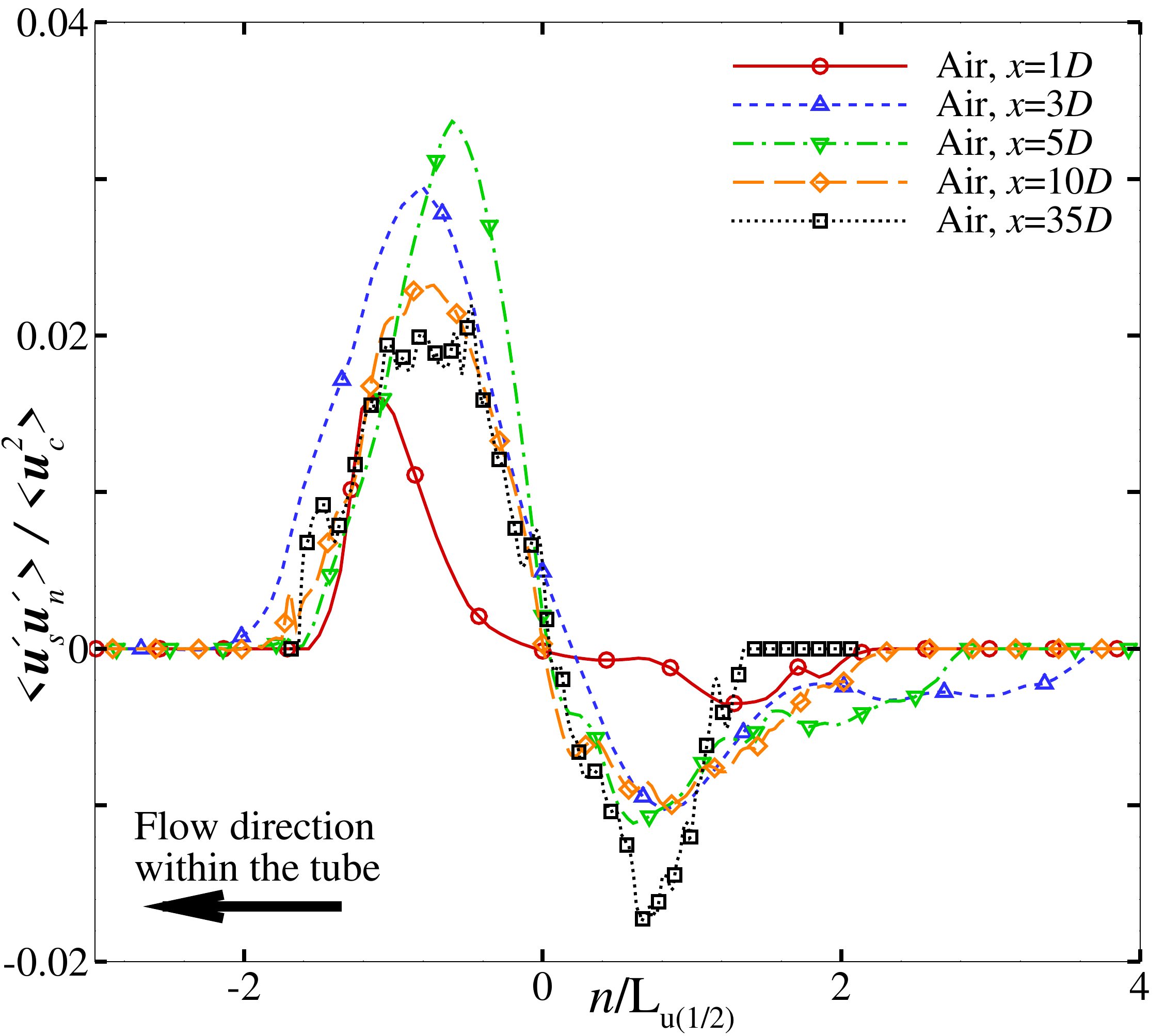} 
	b)\includegraphics[scale=0.27]{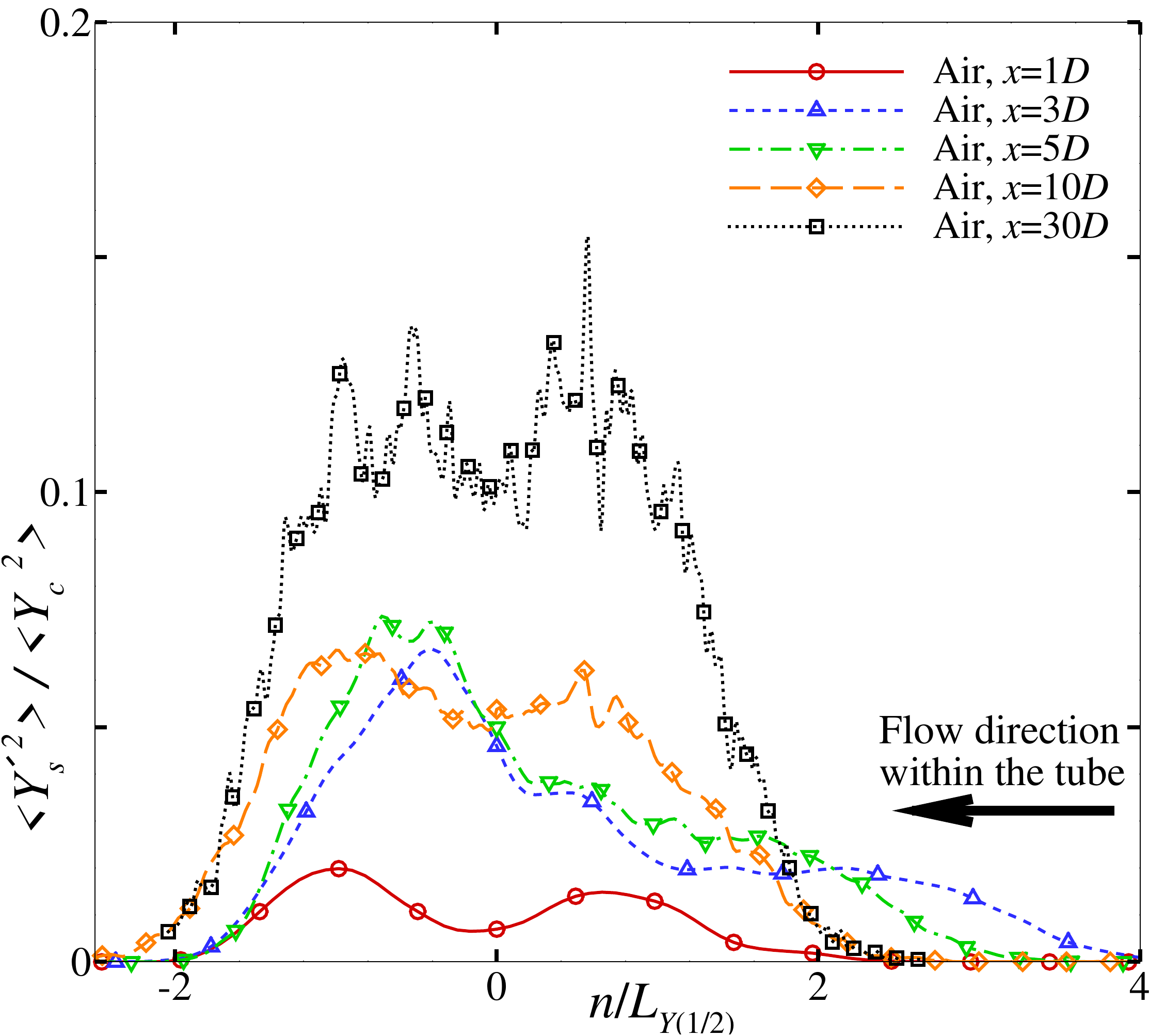}\\
	\raggedright \underline{\textbf{He:}}\\
	\centering
	a)\includegraphics[scale=0.27]{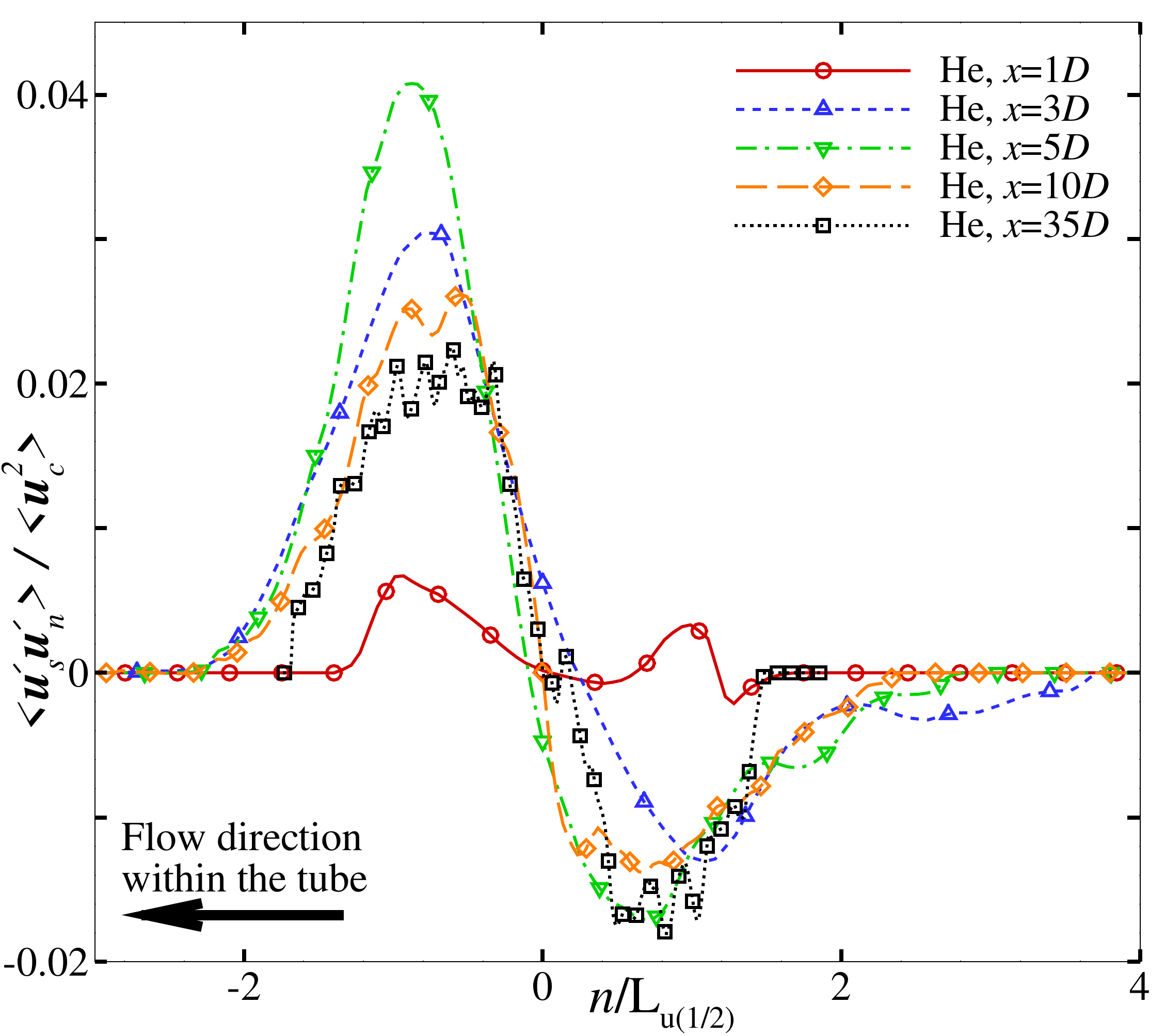} 
	b)\includegraphics[scale=0.27]{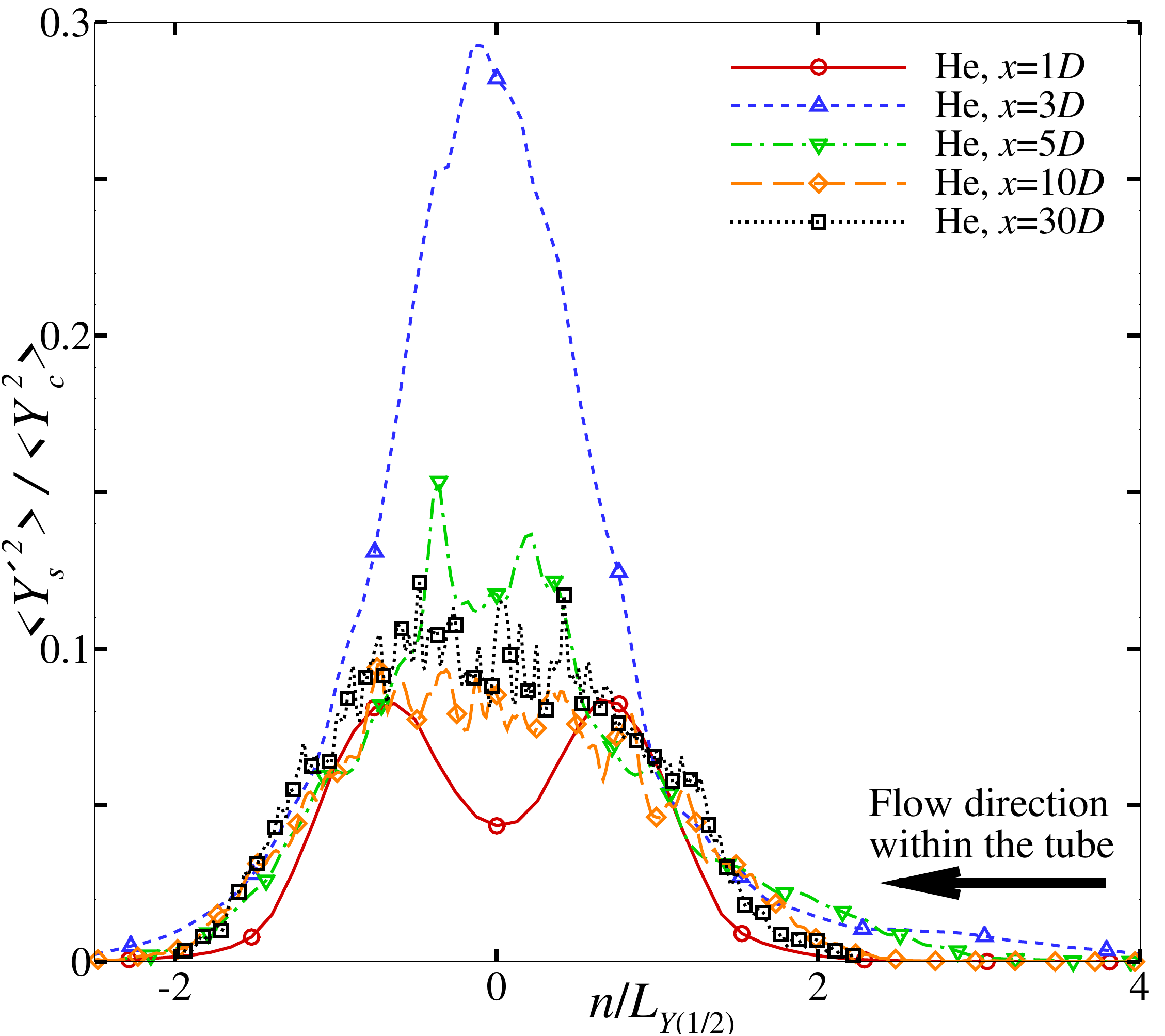}\\
	\caption{a) Normalized time-averaged Reynolds shear stress ($\langle{\boldsymbol{u}^{\prime}_{s} \boldsymbol{u}^{\prime}_{n}}\rangle/ \langle{\boldsymbol{u}^{2}_{c}}\rangle$) and b) concentration variance ($\langle{{Y}^{\prime2}_{s}}\rangle /\langle{{Y}^{2}_{c}}\rangle$) profiles along jet centrelines for air and helium experiments. Here, the profiles are taken at various heights for air and helium measurements in $x$-$z$ planes.}
	\label{fig.Reynolds_Stresses_C_Variances}
\end{figure}     
Higher order statistics were also acquired for the experiments conducted here. The time-averaged Reynolds stress profiles obtained from measurements, $\langle{\boldsymbol{u}^{\prime}_{s} \boldsymbol{u}^{\prime}_{n}}\rangle$, are presented in Fig.\ \ref{fig.Reynolds_Stresses_C_Variances} a). In this case, the Reynolds stress quantities have been normalized by local centreline velocity, $\langle{\boldsymbol{u}^{2}_{c}}\rangle(s)$. In the $x$-$z$ plane, the air and helium experiments captured well the far field self-similar profile, with the helium have slightly higher magnitude of the Reynolds stress compared to the air, as seen before in the axisymmetry jets \citep{Panchapakesan1993JoFM225,Soleimaninia2018IJoHE}. However, to the left of the jet centre (in the $-n$ direction), the horizontal 3D jet experiments were found to have a higher magnitude of the Reynolds stress compared to the axisymmetry jets, in the near field $x\le10D$.  Also, within for $x\le5D$, a higher Reynolds stress was observed beyond $|n/L_{1/2}| < 1$.    

Finally, the normalized concentration variance profiles ($\langle{{Y}^{\prime2}_{s}}\rangle /\langle{{Y}^{2}_{c}}\rangle$), obtained from experiments, are presented in Fig.\ \ref{fig.Reynolds_Stresses_C_Variances} b).  In the $x$-$z$ plane, the initial development of the 3D air jets had a higher variance of concentration to the left of the jet centre (in the $-n$ direction) within the ranges of $x\le10D$. While, helium jet initial profile exhibited semi symmetry saddle-back profile up to $x\le2D$, after this point variances profile recovered the semi Gaussian profile with a maximum magnitude at $x\sim3D$. Beyond $x\ge10D$ jet heights, in the far field, the concentration variance profiles revealed self-similar profile for both air and helium 3D jet experiments.  Also in the core region, much like the axisymmetry jet evolutions, the 3D jet experiments were found to contain a minimum variance near the jet centre, except for helium jet at $x\sim3D$. In general, the magnitudes of mass fraction variance of the helium were higher compared to air, more specifically in the near field.

\section{Discussion}

\subsection{Self-similarity analysis}
In the previous section, for both near and far fields, different velocity and scalar statistical properties were reported for 3D and OP jets of helium and air. It has been well established that these variations are influenced by differences in density, initial conditions and turbulence structures of the jets \citep{Richards1993JoFM417,Xu2002EiF677,Mi2001JoFE878}. Self-similarity (or self-preservation) state in turbulent flows is described as when the flow statistical quantities can be assumed by simple scale factors which depend on only one of the variables. As a consequence, both velocity and scalar pseudo-similarity solutions, in constant or variable density jets, evolve in similar ways when appropriate similarity variables have been used \citep{Panchapakesan1993JoFM225, Pitts1991EiF125, Chen1980}. These pseudo-similarity solutions have been used to develop the analytical models, and approximate the velocity \& scalar decays and growth rates in jet flows. However, It should be noted that the turbulent structure throughout the entire flow field is particularly influenced by the initial jet outflow conditions. As a result, different similarity states in the far field are possible \citep{george1989self, Mi2001JoFM91}. In this section, self-similarity analyses conducted on the current measurement data are presented.

The pseudo-similarity solution, in the turbulent jet, is approximate in the pure jet region, where inertia forces dominate the flow. To estimate the extent of the pure jet region, the following non-dimensional buoyancy length scale (along the $x$-axis, shown in Fig.\ \ref{fig.Experimental_Layout}b) was used \citep{Chen1980}:  
\begin{equation}
x_b=Fr^{-\frac{1}{2}} (\frac{\rho_j}{\rho_{\infty}})^{-\frac{1}{4}} x
\label{eqn.non-dimensional_buoyancy_lengthscale}
\end{equation}
where the Froude number is $Fr$ $(= \frac{\boldsymbol{u}^{2}_{j} \rho_j}{(\rho_{\infty}-\rho_j) gD})$, and $g$ is the acceleration due to gravity. For the flow conditions reported in Table \ref{tab.Flow properties_HSlot1}, $x_b$ varies from 0 to 0.042 for $0<x<30D$, the range of current measurements. Therefore, the hypothetical range of the pure jet region \citep{Chen1980},  $x_b<0.5$, is satisfied for all flow conditions considered in this study. 

The centreline velocity and mass fraction decays for nonreacting jets, for both constant and variable density flows, can be correlated as
\begin{equation}
\frac{\langle{\boldsymbol{u}_{j}}\rangle}{\langle{\boldsymbol{u}_{c}}\rangle} = C_{u} \bigg[\frac{(\textrm{\bf{X}}-\textrm{\bf{X}}_{0,u})}{D_{ef}^*}\bigg]
\label{eqn.Velocity_Decay_Correlation}
\end{equation}
and
\begin{equation}
\frac{\langle{{Y}_{j}}\rangle}{\langle{{Y}_{c}}\rangle} = C_{Y} \bigg[\frac{(\textrm{\bf{X}}-\textrm{\bf{X}}_{0,Y})}{D_{ef}}\bigg]
\label{eqn.Mass_Fraction_Decay_Correlation}
\end{equation}
where the subscripts `$j$' and `$c$' refer to the conditions at the jet exit and centreline, respectively; $\textrm{\bf{X}}_{0,u}$ and $\textrm{\bf{X}}_{0,Y}$ are the dimensional jet virtual origins obtained from inverse centreline velocity and mass fraction decay profiles, respectively, and $C_{u}$ \& $C_{Y}$ are empirical constants obtained with least-mean-square fitting the measured data to Eqs.[\ref{eqn.Velocity_Decay_Correlation}]-[\ref{eqn.Mass_Fraction_Decay_Correlation}].  The concept of effective diameter,
\begin{equation}
	D_{ef}= \frac{2\dot{m}_{j}}{\sqrt{\pi \rho_{\infty} {M}_{j}}}
\label{eqn.effective diameter}
\end{equation}
is defined to account for variations in both the jet fluid density and mean jet exit velocity profile in the turbulent  jet flows \citep{Thring1953SIoC789,Becker1967JoFM285, Dowling1990JoFM109, Pitts1991EiF125, Richards1993JoFM417}; where $\dot{m}_{j}$ and ${M}_{j}$ are the exit mass flux and momentum flux for the jet, respectively. Physically, $D_{ef}$, corresponds to the orifice diameter of a jet having the same momentum and mass flux, but with a density of the ambient fluid instead of the jet fluid. Since asymmetry structures were always observed at the jet exit \citep{Soleimaninia2018IJoHE}, three dimensional measurements of velocity and concentration are required to accurately calculate  $D_{ef}$ in the 3D jets. However, if the density and velocity profiles are uniform at the jet exit, then $D_{ef}$ takes the form as originally introduced by \cite{Thring1953SIoC789}, $D_{ef}=D(\frac{\rho_j}{\rho_{\infty}})^{\frac{1}{2}}$.
It should be noted that in the case of constant density jet (air jet) the effective diameter is equal to the orifice diameter. 

Here, different effective diameter ($D_{ef}$) versions available in the literature, are examined by collapsing the helium data on to the comparable air data, for both hyperbolic decay velocity and scalar laws (Eqs.[\ref{eqn.Velocity_Decay_Correlation}]-[\ref{eqn.Mass_Fraction_Decay_Correlation}]). For the mass fraction decay law, the original effective diameter $D_{ef}=D(\frac{\rho_j}{\rho_{\infty}})^{\frac{1}{2}}$ \citep{Thring1953SIoC789}, used to collapse the scalar data. For the measured velocity data, a modified version of effective diameter, given as $D_{ef}^{**}=D(\frac{\rho_c}{\rho_{\infty}})^{\frac{1}{2}}$ \citep{Talbot2009EiF769}, provides a better correlation in the near field of the flow. Here the subscript `$\infty$' refers to the outer ambient fluid, air. However, the modified version of effective diameter ($D_{ef}^{**}=D(\frac{\rho_c}{\rho_{\infty}})^{\frac{1}{2}}$) requires knowledge of the local centreline concentration; this version cannot be applied in the absence of concentration data . Upon further analysis, it was found that if the second root, in the original effective diameter ($D_{ef}=D(\frac{\rho_j}{\rho_{\infty}})^{\frac{1}{2}}$), is replaced by $\sim$ thrid root, then the velocity data shows good correlation with the collapsed curves of the aggregate 3D jet data  in both the near and far fields. Therefore, this new modified version of  effective diameter, $D_{ef}^*=D(\frac{\rho_j}{\rho_{\infty}})^{0.3}$, was used to correlate the centreline velocity decay (Eq.[\ref{eqn.Velocity_Decay_Correlation}]). It should be noted that the latter modified version of effective diameter, $D_{ef}^*$, may only valid for the current 3D jet experiments, due to the effects of specific conditions at the jet such as geometry, effective surface area, flow structures, density profiles, and velocity profiles.

\begin{figure}
	\centering
	a)\includegraphics[scale=0.27]{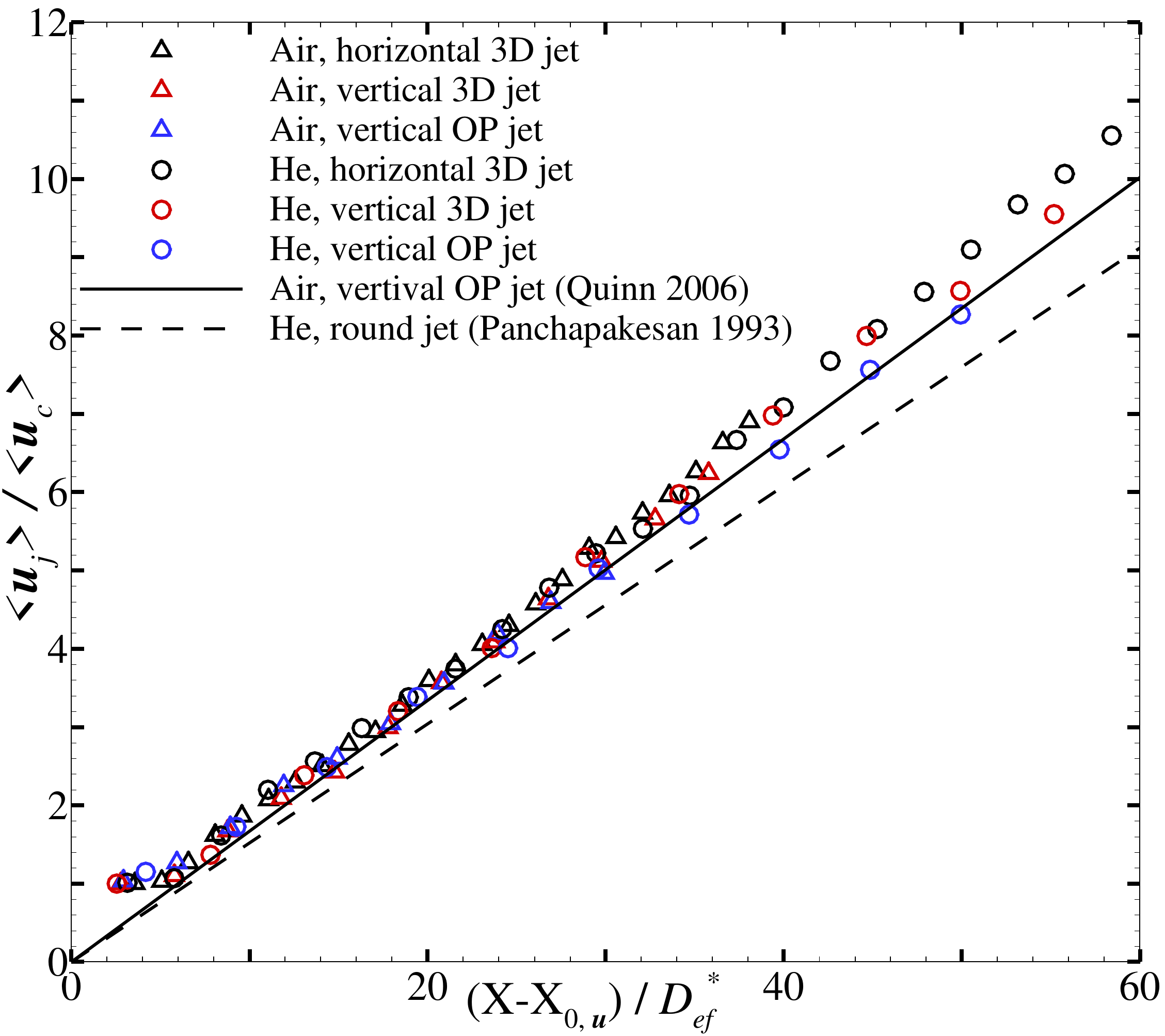} 
	b)\includegraphics[scale=0.27]{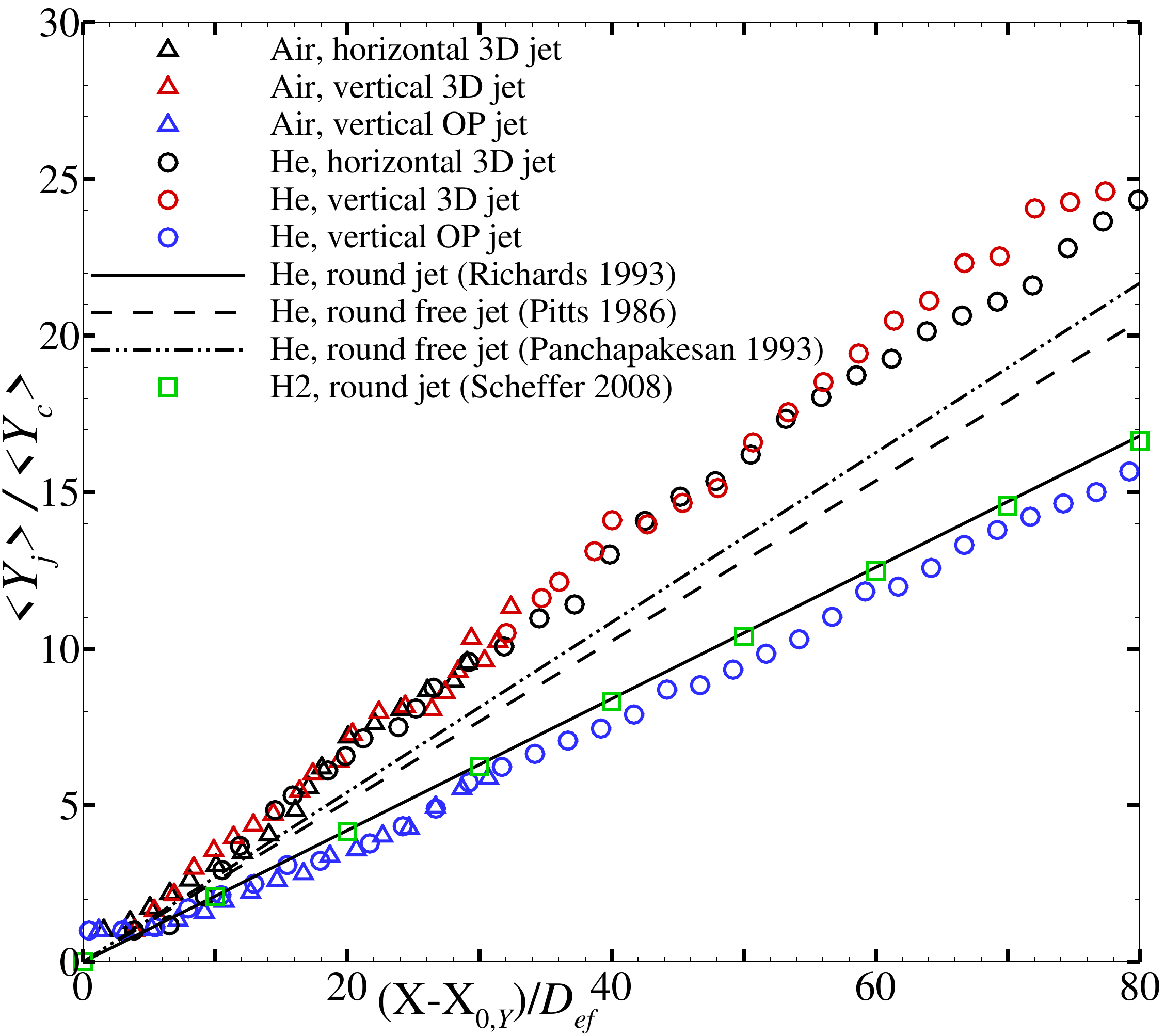}\\
	\caption{Inverse axial velocity and mass fraction decay along jet centrelines versus downstream distance non-dimensionalized by $D_{ef}^*$ and $D_{ef}$, a) velocity ($\langle{\boldsymbol{u}_{j}}\rangle/ \langle{\boldsymbol{u}_{c}}\rangle$) and b) mass fraction ($\langle{{Y}_{j}}\rangle/ \langle{{Y}_{c}}\rangle$) for experiments. Also shown, for comparison {are} vertical 3D \& OP jets, and round pipe He \& H2 jet experiments \citep{Soleimaninia2018IJoHE, Richards1993JoFM417, Schefer2008IJoHE6373}.}
	\label{fig.Velocity_MassFraction_Decay_Reverse_Similarity}
\end{figure} 

In Fig.\ \ref{fig.Velocity_MassFraction_Decay_Reverse_Similarity}, the centreline revolution of the inverse velocity (Fig.\ \ref{fig.Velocity_JetDecay_SpreadingRate} a) and mass fraction (Fig.\ \ref{fig.Concentration_JetDecay_SpreadingRate} a) profiles have been reconstructed for the 3D jets as a function of distance from the virtual origins, normalized with effective diameter, i.e. $[{(\textrm{\bf{X}}-\textrm{\bf{X}}_{0})}/{D_{ef}}]$. Self-similarity decay lines, obtained by curve fitting the results, are also shown for OP, Smooth Contraction (SC), pipe round free, and pipe round confined jets \citep{Quinn2006EJoM-B279,Panchapakesan1993JoFM225,Richards1993JoFM417,Pitts1986,Schefer2008IJoHE6373}. The experimental data for all helium jets were collapsed onto the comparable air jets, verifying that the correct version of effective diameter along with virtual origin distances are the appropriate scaling parameters to correlate both velocity and scalar pseudo-similarity solutions in the constant or variable density jets. The velocity decay rates of all 3D jets are very similar to OP jets based on a comparison of velocity decay plots (\ref{fig.Velocity_MassFraction_Decay_Reverse_Similarity} a). However, the mass fraction decay profiles of 3D jets show a steeper decay rate than is observed for OP jets (\ref{fig.Velocity_MassFraction_Decay_Reverse_Similarity} b). This observation further supports the fact that the velocity field spreads slower than the concentration field. This conclusion is supported by the preferential transport of scalar quantities over momentum flux that is evident in previous studies \citep{Talbot2009EiF769,Lubbers2001FDR189}. It is also clear that a pipe confined jet of hydrogen \citep{Schefer2008IJoHE6373} follows the same decay rate of those pipe confined jets of helium \citep{Richards1993JoFM417}. This observation is consistent with the fact that the scalar decay rate is independent of initial density ratio but influenced notably by the jet initial conditions and other potential factors such as measurement conditions.
\begin{table}
  \begin{center}
  	\def~{\hphantom{0}}
		\begin{tabular}{lccccc}
			Jet &  $R_{\rho}$ & $\textrm{\bf{X}}_{0,u}/D$ &$C_{\boldsymbol{u}}$  &$\textrm{\bf{X}}_{0,Y}/D$  & $C_{Y}$\\[3pt]						
			Air, 3D Horizontal&1&-3.07&0.174  & -1.07 & 0.326\\
			Air, 3D Vertical&1&-2.78&0.170 & -3.39 & 0.319\\
			Air, OP Vertical&1&3.08&0.169 & -0.68 & 0.224\\			
			Air, OP Vertical \citep{Quinn2006EJoM-B279}&1&2.15&0.167 & -- & --\\			
			He, 3D Horizontal&0.14&-1.29&0.175 & -0.95 & 0.313\\			
			He, 3D Vertical&0.14&-1.45&0.170 & -6.42 & 0.316\\			
			He, OP Vertical&0.14&3.54&0.170 & 2.32 & 0.221\\			
			He, SC Vertical \citep{Panchapakesan1993JoFM225}&0.14&--&0.152 & -- & 0.271\\			
			He, Pipe Vertical \citep{Richards1993JoFM417}&0.14&--&-- & 3.0 & 0.212\\			
			He, Pipe Vertical \citep{Pitts1986}&0.14&--&-- & 4.45 & 0.256 \\			
			H2, Pipe Vertical \citep{Schefer2008IJoHE6373}&0.069&--&-- & 4.0 & 0.208 \\			
	\end{tabular}
	\caption{Centerline velocity and scalar pseudo-similarity decay properties}
	\label{tab.Centerline_Decay_Properties}
 \end{center}
\end{table}
Further comparisons of the centreline pseudo-similarity decay properties are shown in Table \ref{tab.Centerline_Decay_Properties}. Here, $R_{\rho}$ is a ratio of the jet fluid density to the ambient fluid, $R_{\rho} = \frac{\rho_j}{\rho_{\infty}}$. For both velocity and mass fraction quantities, these self-similarity properties are obtained from data fitted by a least-mean-square algorithm to Eqs. (\ref{eqn.Velocity_Decay_Correlation})-(\ref{eqn.Mass_Fraction_Decay_Correlation}). Table \ref{tab.Centerline_Decay_Properties} also provides a comparison of self-similarity properties of OP, SC, pipe round free, and pipe round confined jets \citep{Quinn2006EJoM-B279,Panchapakesan1993JoFM225,Richards1993JoFM417,Pitts1986,Schefer2008IJoHE6373}. It should be noted that dimensional virtual origin distances, $\textrm{\bf{X}}_{0,u}$ and $\textrm{\bf{X}}_{0,Y}$, are normalized by the orifice diameter ($D$).

Upon comparison of the velocity decay slopes, for the air OP jets, the value of $C_u=0.169$ is in good agreement with the value of 0.167 reported previously for the air OP jet \citep{Quinn2006EJoM-B279}. The small difference is associated with higher Reynolds number of $Re = 1.84\times10^5$ compared to present study ($Re = 1.65\times10^4$), which results in a decrease of the velocity decay rate. The helium OP jet shows slightly higher decay rate to that of the air OP jet, as shown previously in Fig.\ \ref{fig.Velocity_JetDecay_SpreadingRate} a, but with a minor increase of the virtual origin, $x_{0u}$. It is well known that the virtual origin of a jet is highly influenced by the jet initial conditions and does not vary in any systematic manner. The vertical helium and air 3D jets have almost the same decay slopes, whereas the horizontal helium 3D jet has a slightly higher slope than the comparable horizontal air jet, as previously noticed in Fig.\ \ref{fig.Velocity_JetDecay_SpreadingRate} a. In general, a higher velocity decay rate is observed for 3D jets compared to OP, SC and pipe jets based on a comparison of the velocity decay slopes. This is associated with enhanced turbulent mixing in the 3D jets, as a result of their asymmetry flow pattern, specifically in the near field. 

In contrast, by comparing the helium mass fraction decay slopes in table \ref{tab.Centerline_Decay_Properties}, it is found that reported $C_Y$ values in the literature for SC and pipe jets are larger than those OP values obtained in the current study. This is in contrast with the well established fact that the OP jets exhibit the highest mixing rate, followed by the SC jets and finally the pipe jets \citep{Mi2001JoFE878}. It should be noted that the value of $C_Y=0.271$ reported for SC helium jet \citep{Panchapakesan1993JoFM225}, is obtained without considering the scalar virtual origin, $\textrm{\bf{X}}_{0,Y}$, in Eq.(\ref{eqn.Mass_Fraction_Decay_Correlation}). In addition, $C_Y=0.256$  reported for the pipe jet \citep{Pitts1986}, is correlated based on a different exponent in the effective diameter equation. The pipe jet data has been correlated using the factor of $(\frac{\rho_j}{\rho_{\infty}})^{0.6}$ instead of $(\frac{\rho_j}{\rho_{\infty}})^{\frac{1}{2}}$ in the original version of $D_{ef}$ which would explain the higher $C_Y$ value reported for the pipe jet in their measurements. The mass fraction decay slope observed for helium 3D jets is smaller than for air 3D jets. However, upon comparison of mass fraction decay slopes between the 3D and other jets, it is found that the 3D jets have the highest decay slopes.  This result further supports the fact that significantly higher turbulent mixing and entertainment rates occur in the 3D jets compared to the axisymmetry jets, as recently concluded in the experimental and numerical study on vertical 3D jet \citep{Soleimaninia2018IJoHE}.

\subsection{Buoyancy effect}
For the 3D jets, it was found that the perpendicular stream-wise axis of the hole, relative to the flow direction within the tube, resulted into a deflection of the jet away from it's horizontal $x-$axis. Initially, from Fig. \ref{fig.Jet_Centerline_Trajectory}, all 3D jets emerged with a similar deflection angle. But only after $x>2D$, both horizontal and vertical air jets were found to deflect more than helium jets. Buoyancy forces, aside from less significant contributors, are a probable cause of the increased deflection observed for vertical air jets in comparison to helium jets. But from the comparison of helium jets centreline trajectories (Fig. \ref{fig.Jet_Centerline_Trajectory}), buoyancy effect is clearly the main contributor in the significant deflection of horizontal case from the horizontal $x-$axis compared to vertical jet. Whereas such deflection were not observed in horizontal air jet compared to vertical case. 

\begin{figure}
	\centering
	\includegraphics[scale=0.35]{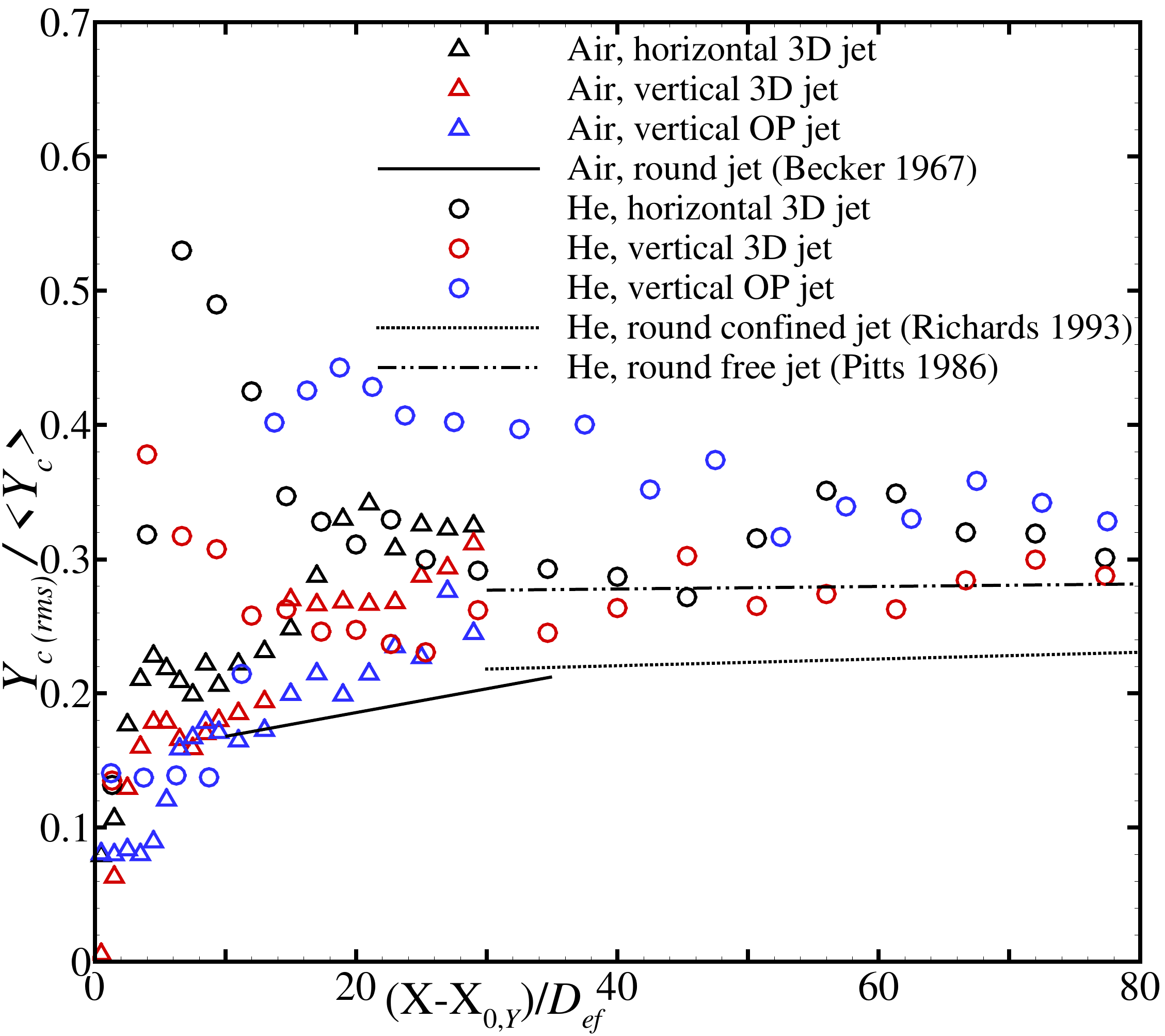} \\
	\caption{Centreline evolution of normalized mass fraction fluctuation intensities, ${Y}_{c(rms)}/ \langle{{Y}_{c}}\rangle$, versus downstream distance non-dimensionalized by $D_{ef}$ for experiments. Also shown, for comparison are vertical 3D \& OP jets, and round pipe jet experiments \citep{Soleimaninia2018IJoHE,Becker1967JoFM285,Pitts1991EiF125,Richards1993JoFM417}.}
	\label{fig.Concenration_fluctuation_rms_Similarity}
\end{figure} 

Figure \ref{fig.Concenration_fluctuation_rms_Similarity} reconstructed the unmixedness profile (Fig. \ref{fig.Concenration_fluctuation_rms}) for the 3D jets as a function of distances from the virtual origin ($\textrm{\bf{X}}_{0,Y}$) and normalized by effective diameter ($D_{ef}$). Along with same remarks observed as those presented in Fig. \ref{fig.Concenration_fluctuation_rms}, it is clear that effective diameter would not be a necessary length scale for unmixedness profile, since helium and air data are already collapsed on the same curves by scaling with the jet orifice diameter ($D$). All 3D jets recovered the asymptotic value of $\sim26\%$ , reported for variable density free round jet \citep{Pitts1986}, which is closer to the orifice compared to axisymmetry OP jets. Further downstream, the horizontal 3D jets reached a higher asymptotic plateau ($\sim33\%$) in the far field. This difference might be solely associated with buoyancy, which becomes dominant closer to the orifice, in the horizontal cases compared to the vertical 3D jet. Other parameters such as co-flow, initial conditions, Reynolds number, and measurement uncertainty could also play a significant role \citep{Pitts1991EiF125}. However, their effects are  negligible since the similar experimental system and parameters have been used in current measurements. Despite these differences, it is clear that centreline unmixedness is independent of $R_{\rho}$ and achieves asymptotic value based on the initial conditions at some downstream distance, influenced by Reynolds number. However, the initial increase in the mass fraction fluctuation intensity in the near field occurs more rapidly in lower density gas, helium compared to air.

\subsection{Asymmetry effect}
For 3D jets, flow separation of the emerging gas originating from inside the tube, similar to flow over a backward step, is always observed at the entrance of orifice. This phenomenon was also previously reported in vertical 3D jets \citep{Soleimaninia2018IJoHE}. This flow separation contributes to the velocity and scalar deficit near the edge of the orifice located on the lower side of the jet (in the $+n$ direction), and results in a slight contraction in the width of the jet has been observed in both velocity and concentration field (Figs. \ref{fig.Velocity_JetDecay_SpreadingRate}b \& \ref{fig.Concentration_JetDecay_SpreadingRate}b) in the range of $1D<x<4D$. Asymmetry structure was always observed for all 3D jet, owing to flow separation and associated deficits in velocity and scalar field. This asymmetry pattern is clearly evident in the lower side of the jet centrelines (in the $+n$ direction), where more velocity and mass fraction spreading is exhibited near the tail ends of the radial profiles,  $1<(n/L_{1/2})$ for $x\le5D$ (Fig. \ref{fig.Vs_Cs}). This asymmetry pattern and three-dimension nature of the 3D jets, encouraged more entrainment which lead to enhanced turbulent mixing in the 3D jets compared to the axisymmetry jets. This enhancement is clearly observed, upon comparison between the 3D and axisymmetry jets, in the velocity and mass fraction decays and spreading rates, and their pseudo-similarity solution presented in Figs. \ref{fig.Velocity_JetDecay_SpreadingRate}, \ref{fig.Concentration_JetDecay_SpreadingRate} \& \ref{fig.Velocity_MassFraction_Decay_Reverse_Similarity}, respectively.

Much like 3D jets, non-circular jets are also well-known to entrain ambient fluid more effectively than their axisymmetry round jets counterparts \citep{Gutmark1999ARoFM239}. The enhanced mixing in non-circular jet is associated with a higher degree of three-dimensionality in the coherent structures of the flow. As the jet spreads, the deformation dynamics of asymmetric vortices yields a complex topology, which results in the interaction of streamwise and azimuthal vortices and the associated energy transfer between them. This phenomena, ``axis-switching'', is the main fundamental mechanism for the enhanced entrainment properties of the non-circular jets, and it has been only reported in the non-circular jets \citep{Gutmark1999ARoFM239,Mi2010FTaC583}. Generally, the axis-switching can be observed from cross passing the jet half-width profiles in the major and minor axis planes ($s$-$n$ and $s$-$y$ planes). This phenomenon also can be observed from the broad humps in axial development of the tangential turbulence intensity along the jet centrelines ($\boldsymbol{u}_{s(rms)}/ \langle{\boldsymbol{u}_{c}}\rangle$) as concluded in comparative experimental study on the non-circular jets \citep{Mi2010FTaC583}. In Fig.\ \ref{fig.Velocity_fluctuation_rms}a,  while no humps occurs in the variation of ($\boldsymbol{u}_{s(rms)}/ \langle{\boldsymbol{u}_{c}}\rangle$) for the round OP or pipe jets, humps are present for the round 3D jets. This can be correlated with the axis-switching phenomenon, and as a result of enhanced entrainment, also correlates with increased centerline velocity decay rates.  This, in turn, results in higher values of ($\boldsymbol{u}_{s(rms)}/ \langle{\boldsymbol{u}_{c}}\rangle$). This phenomenon is in fact observed in the recent study on vertical 3D jet \citep{Soleimaninia2018IJoHE}, where the air jet half-width profiles cross-pass at approximately $15D$ from the orifice. Therefore, axis-switching would be one of the main underlying phenomena responsible for enhanced turbulent mixing and entrainment of the 3D jets. 

\section{Conclusions}
In this study, simultaneous velocity and concentration measurements were conducted in order to investigate horizontal turbulent jets, of varying gas densities and Reynolds numbers, issuing from a round orifice machined on the side of a round tube. The fluids considered were air and helium. The results were compared to previous studies of vertical jet, issuing from the same pipeline geometry and axisymmetric round OP jets \citep{Soleimaninia2018IJoHE}.  Comparisons were also made with horizontal axisymmetric round jets, issuing through flat plates \citep{Ash2012}, and the results of other relevant experimental studies of constant and variable density turbulent axisymmetry jets. 

By considering flow emerging through a hole located on the side of a tube wall, it was found that the flow arrangement caused a significant deflection from the axis normal to the orifice. This characteristic was also previously observed in vertical jets of the similar pipeline configuration \citep{Soleimaninia2018IJoHE}.  In the current investigation, the helium jet deflection was found to be initially governed by the density of the gas in the near field, and it experienced further deflection due to buoyancy in the far field.  The buoyancy-caused deviation in the far field  was found to be well reproduced by a power law expression with the exponent $\sim1.3$.  In contrast, it was found that such buoyancy effects were not present in axisymmetric round jet helium experiments, where the jet issued through flat-plates, with the same Froude number.  This observation suggests that  the realistic leak geometry along the pipeline orientation considered in this study causes buoyancy effects to dominate much closer to the orifice than expected for axisymmetric round jets.  Furthermore, it was found that buoyancy effects have  a negligible impact on the decay of jet velocity and spreading rates. This implication is also true for fluctuation quantities, where buoyancy was found to not have a significant effect on centreline velocity fluctuation intensities. Nevertheless, higher centreline mass fraction fluctuation intensities ($\sim33\%$) for the horizontal 3D jets compared to the vertical cases ($\sim26\%$), may have been caused by buoyancy effect. 

Owing to asymmetry flow structure and three-dimension nature of 3D jets, enhanced turbulent mixing was always observed in 3D jets compared to axisymmetry jets. This enhanced mixing and entrainment caused the reduction in the potential-core length, as well as increases in the decay and spreading rates of both velocity and concentration. Despite the fact that the orifice geometry is round, the axis-switching phenomenon was observed in 3D jets, and is believed to be one of the main fundamental mechanisms for the enhanced entrainment properties of the 3D jets. Furthermore, the 3D jets obey the pseudo-similarity decay law with the scaling indicated by the effective diameter. The mass fraction decay along the centreline scaled well with the original effective diameter term ($D_{ef}=D(\frac{\rho_j}{\rho_{\infty}})^{\frac{1}{2}}$), while the modified version of the effective diameter ($D_{ef}^*=D(\frac{\rho_j}{\rho_{\infty}})^{0.3}$) provided a more accurate velocity decay profile. Finally, it was shown that the turbulent velocity and scalar properties are dependent on the initial jet conditions for all regions of the flow field. Therefore, caution is required when using round axisymmetry jet assumptions to correlate the correct dispersion, velocity and concentration decay rates and, consequently, the extent of flammability envelope of a jet emitted from realistic leak geometries.

The authors would like to acknowledge funding from the Natural Sciences and Engineering Research Council of Canada (NSERC).

\bibliographystyle{jfm}
\bibliography{Literature_Review}

\begin{thebibliography}{46}
\expandafter\ifx\csname natexlab\endcsname\relax\def\natexlab#1{#1}\fi
\def\au#1{#1} \def\ed#1{#1} \def\yr#1{#1}\def\at#1{#1}\def\jt#1{\textit{#1}}
  \def\bt#1{#1}\def\bvol#1{\textbf{#1}} \def\vol#1{#1} \def\pg#1{#1}
  \def\publ#1{#1}\def\arxiv#1{#1}\def\org#1{#1}\def\st#1{\textit{#1}}

\bibitem[Amielh {\em et~al.\/}(1996)Amielh, Djeridane, Anselmet \&
  Fulachier]{Amielh1996IJoHaMT2149}
{\sc \au{Amielh, M.}, \au{Djeridane, T.}, \au{Anselmet, F.} \& \au{Fulachier,
  L.}} \yr{1996}  \at{Velocity near-field of variable density turbulent jets}.
  \jt{International Journal of Heat and Mass Transfer}  \bvol{39}~(10),
  \pg{2149--2164}.

\bibitem[Ash(2012)]{Ash2012}
{\sc \au{Ash, A.}} \yr{2012} Quantitative imaging of multi-component turbulent
  jets. Master's thesis, University of Victoria.

\bibitem[Ball {\em et~al.\/}(2012)Ball, Fellouah \& Pollard]{Ball2012PiAS1}
{\sc \au{Ball, C.G.}, \au{Fellouah, H.} \& \au{Pollard, A.}} \yr{2012}  \at{The
  flow field in turbulent round free jets}.  \jt{Progress in Aerospace
  Sciences}  \bvol{50},  \pg{1 -- 26}.

\bibitem[Becker {\em et~al.\/}(1967)Becker, Hottel \&
  Williams]{Becker1967JoFM285}
{\sc \au{Becker, H.~A.}, \au{Hottel, H.~C.} \& \au{Williams, G.~C.}} \yr{1967}
  \at{The nozzle-fluid concentration field of the round, turbulent , free jet}.
   \jt{Journal of Fluid Mechanics}  \bvol{30},  \pg{285--303}.

\bibitem[Carazzo {\em et~al.\/}(2006)Carazzo, Kaminski \&
  Tait]{Carazzo2006JoFM137}
{\sc \au{Carazzo, G.}, \au{Kaminski, E.} \& \au{Tait, S.}} \yr{2006}  \at{The
  route to self-similarity in turbulent jets and plumes}.  \jt{Journal of Fluid
  Mechanics}  \bvol{547},  \pg{137--148}.

\bibitem[Chen \& Rodi(1980)]{Chen1980}
{\sc \au{Chen, C.~J.} \& \au{Rodi, W.}} \yr{1980} {\em Vertical turbulent
  buoyant jets : a review of experimental data\/}.  \publ{Oxford; New York:
  Pergamon Press}.

\bibitem[Chernyavsky {\em et~al.\/}(2011)Chernyavsky, Wu, P\`eneau, B\`enard,
  Oshkai \& Djilali]{Chernyavsky2011IJoHE2645}
{\sc \au{Chernyavsky, B.}, \au{Wu, T.~C.}, \au{P\`eneau, F.}, \au{B\`enard,
  P.}, \au{Oshkai, P.} \& \au{Djilali, N.}} \yr{2011}  \at{Numerical and
  experimental investigation of buoyant gas release: Application to hydrogen
  jets}.  \jt{International Journal of Hydrogen Energy}  \bvol{36}~(3),
  \pg{2645--2655}.

\bibitem[Dowling \& Dimotakis(1990)]{Dowling1990JoFM109}
{\sc \au{Dowling, D.~R.} \& \au{Dimotakis, P.~E.}} \yr{1990}  \at{Similarity of
  the concentration field of gas-phase turbulent jets}.  \jt{Journal of Fluid
  Mechanics}  \bvol{218},  \pg{109}.

\bibitem[Ekoto {\em et~al.\/}(2012)Ekoto, Houf, Evans, Merilo \&
  Groethe]{Ekoto2012IJoHE17446}
{\sc \au{Ekoto, I.~W.}, \au{Houf, W.~G.}, \au{Evans, G.~H.}, \au{Merilo, E.~G.}
  \& \au{Groethe, M.~A.}} \yr{2012}  \at{Experimental investigation of hydrogen
  release and ignition from fuel cell powered forklifts in enclosed spaces}.
  \jt{International Journal of Hydrogen Energy}  \bvol{37}~(22),  \pg{17446 --
  17456}, hySafe 1.

\bibitem[George(1989)]{george1989self}
{\sc \au{George, W.~K.}} \yr{1989}  \at{The self-preservation of turbulent
  flows and its relation to initial conditions and coherent structures}.
  \jt{Advances in turbulence (Arndt, R. E. A. \& George, W. K., Eds.)}  \pg{pp.
  39--73}.

\bibitem[Gutmark \& Grinstein(1999)]{Gutmark1999ARoFM239}
{\sc \au{Gutmark, E.~J.} \& \au{Grinstein, F.~F.}} \yr{1999}  \at{Flow control
  with noncircular jets}.  \jt{Annual Review of Fluid Mechanics}
  \bvol{31}~(1),  \pg{239--272},  \arxiv{arXiv:
  http://dx.doi.org/10.1146/annurev.fluid.31.1.239}.

\bibitem[Hajji {\em et~al.\/}(2015)Hajji, Jouini, Bouteraa, Elcafsi, Belghith
  \& Bournot]{Hajji2015IJoHE9747}
{\sc \au{Hajji, Y.}, \au{Jouini, B.}, \au{Bouteraa, M.}, \au{Elcafsi, A.},
  \au{Belghith, A.} \& \au{Bournot, P.}} \yr{2015}  \at{Numerical study of
  hydrogen release accidents in a residential garage}.  \jt{International
  Journal of Hydrogen Energy}  \bvol{40}~(31),  \pg{9747 -- 9759}.

\bibitem[Houf {\em et~al.\/}(2013)Houf, Evans, Ekoto, Merilo \&
  Groethe]{Houf2013IJoHE8179}
{\sc \au{Houf, W.G.}, \au{Evans, G.H.}, \au{Ekoto, I.W.}, \au{Merilo, E.G.} \&
  \au{Groethe, M.A.}} \yr{2013}  \at{Hydrogen fuel-cell forklift vehicle
  releases in enclosed spaces}.  \jt{International Journal of Hydrogen Energy}
  \bvol{38}~(19),  \pg{8179 -- 8189}.

\bibitem[Houf \& Schefer(2008)]{Houf2008IJoHE1435}
{\sc \au{Houf, W.} \& \au{Schefer, R.}} \yr{2008}  \at{Analytical and
  experimental investigation of small-scale unintended releases of hydrogen}.
  \jt{International Journal of Hydrogen Energy}  \bvol{33}~(4),  \pg{1435 --
  1444}.

\bibitem[Houf {\em et~al.\/}(2010)Houf, Schefer, Evans, Merilo \&
  Groethe]{Houf2010IJoHE4758}
{\sc \au{Houf, W.}, \au{Schefer, R.}, \au{Evans, G.}, \au{Merilo, E.} \&
  \au{Groethe, M.}} \yr{2010}  \at{Evaluation of barrier walls for mitigation
  of unintended releases of hydrogen}.  \jt{International Journal of Hydrogen
  Energy}  \bvol{35}~(10),  \pg{4758 -- 4775}, novel Hydrogen Production
  Technologies and Applications.

\bibitem[Hussein {\em et~al.\/}(1994)Hussein, Capp \&
  George]{Hussein1994JoFM31}
{\sc \au{Hussein, H.~J.}, \au{Capp, S.~P.} \& \au{George, W.~K.}} \yr{1994}
  \at{Velocity measurements in a high-reynolds-number, momentum-conserving,
  axisymmetric, turbulent jet}.  \jt{Journal of Fluid Mechanics}  \bvol{258},
  \pg{31--75}.

\bibitem[Iverson {\em et~al.\/}(2015)Iverson, DeVaal, Kerr \&
  Oshkai]{Iverson2015IJoHE13134}
{\sc \au{Iverson, D.}, \au{DeVaal, J.}, \au{Kerr, J.} \& \au{Oshkai, P.}}
  \yr{2015}  \at{Investigation of ignited hydrogen leaks from tube fittings}.
  \jt{International Journal of Hydrogen Energy}  \bvol{40}~(38),  \pg{13134 --
  13145}.

\bibitem[Lewis \& {von Elbe}(1961)]{Lewis1961}
{\sc \au{Lewis, B.} \& \au{{von Elbe}, G.}} \yr{1961} {\em Combustion, Flames
  and Explosions of Gases\/}, 2nd edn.  \publ{New York: Academic Press}.

\bibitem[Lipari \& Stansby(2011)]{Lipari2011FTaC79}
{\sc \au{Lipari, G.} \& \au{Stansby, P.~K.}} \yr{2011}  \at{Review of
  experimental data on incompressible turbulent round jets}.  \jt{Flow,
  Turbulence and Combustion}  \bvol{87}~(1),  \pg{79--114}.

\bibitem[Lubbers {\em et~al.\/}(2001)Lubbers, Brethouwer \&
  Boersma]{Lubbers2001FDR189}
{\sc \au{Lubbers, C.~L.}, \au{Brethouwer, G.} \& \au{Boersma, B.~J.}} \yr{2001}
   \at{Simulation of the mixing of a passive scalar in a round turbulent jet}.
  \jt{Fluid Dynamics Research}  \bvol{28}~(3),  \pg{189}.

\bibitem[M.~Kaushik(2015)]{Kaushik2015AJoFD1}
{\sc \au{M.~Kaushik, R.~Kumar, Humrutha~G.}} \yr{2015}  \at{Review of
  computational fluid dynamics studies on jets}.  \jt{American Journal of Fluid
  Dynamic}  \bvol{5}~(A),  \pg{1--11}.

\bibitem[Maxwell {\em et~al.\/}(2017)Maxwell, Soleimani~nia, Oshkai \&
  Djilali]{Maxwell2017}
{\sc \au{Maxwell, B.}, \au{Soleimani~nia, M.}, \au{Oshkai, P.} \& \au{Djilali,
  N.}} \yr{2017} Large eddy simulations of asymmetric turbulent hydrogen jets
  issuing from realistic pipe geometries.  \bt{In {\em Proceedings of the 7th
  International Conference on Hydrogen Safety, Hamburg, Germany\/}}.

\bibitem[Mi {\em et~al.\/}(2007)Mi, Kalt, Nathan \& Wong]{Mi2007EiF625}
{\sc \au{Mi, J.}, \au{Kalt, P.}, \au{Nathan, G.~J.} \& \au{Wong, C.~Y.}}
  \yr{2007}  \at{Piv measurements of a turbulent jet issuing from round
  sharp-edged plate}.  \jt{Experiments in Fluids}  \bvol{42}~(4),
  \pg{625--637}.

\bibitem[Mi \& Nathan(2010)]{Mi2010FTaC583}
{\sc \au{Mi, J.} \& \au{Nathan, G.~J.}} \yr{2010}  \at{Statistical properties
  of turbulent free jets issuing from nine differently-shaped nozzles}.
  \jt{Flow, Turbulence and Combustion}  \bvol{84}~(4),  \pg{583--606}.

\bibitem[Mi {\em et~al.\/}(2001{\natexlab{{\em a\/}}})Mi, Nathan \&
  Nobes]{Mi2001JoFE878}
{\sc \au{Mi, J.}, \au{Nathan, G.~J.} \& \au{Nobes, D.~S.}}
  \yr{2001{\natexlab{{\em a\/}}}}  \at{Mixing characteristics of axisymmetric
  free jets from a contoured nozzle, an orifice plate and a pipe}.  \jt{Journal
  of Fluids Engineering}  \bvol{123}~(4),  \pg{878--883}.

\bibitem[Mi {\em et~al.\/}(2001{\natexlab{{\em b\/}}})Mi, Nobes \&
  Nathan]{Mi2001JoFM91}
{\sc \au{Mi, J.}, \au{Nobes, D.~S.} \& \au{Nathan, G.~J.}}
  \yr{2001{\natexlab{{\em b\/}}}}  \at{Influence of jet exit conditions on the
  passive scalar field of an axisymmetric free jet}.  \jt{Journal of Fluid
  Mechanics}  \bvol{432},  \pg{91--125}.

\bibitem[Panchapakesan \& Lumley(1993)]{Panchapakesan1993JoFM225}
{\sc \au{Panchapakesan, N.~R.} \& \au{Lumley, J.~L.}} \yr{1993}  \at{Turbulence
  measurements in axisymmetric jets of air and helium. part 2. helium jet}.
  \jt{Journal of Fluid Mechanics}  \bvol{246},  \pg{225--247}.

\bibitem[Pitts(1986)]{Pitts1986}
{\sc \au{Pitts, W.~M.}} \yr{1986} {\em Effects of global density and Reynolds
  number variations on mixing in turbulent, axisymmetric jets\/}.
  \publ{National Bureau of Standards, US Department of Commerce}.

\bibitem[Pitts(1991{\natexlab{{\em a\/}}})]{Pitts1991EiF125}
{\sc \au{Pitts, W.~M.}} \yr{1991{\natexlab{{\em a\/}}}}  \at{Effects of global
  density ratio on the centerline mixing behavior of axisymmetric turbulent
  jets}.  \jt{Experiments in Fluids}  \bvol{11}~(2),  \pg{125--134}.

\bibitem[Pitts(1991{\natexlab{{\em b\/}}})]{Pitts1991EiF135}
{\sc \au{Pitts, W.~M.}} \yr{1991{\natexlab{{\em b\/}}}}  \at{Reynolds number
  effects on the mixing behavior of axisymmetric turbulent jets}.
  \jt{Experiments in Fluids}  \bvol{11}~(2),  \pg{135--141}.

\bibitem[Quinn(2006)]{Quinn2006EJoM-B279}
{\sc \au{Quinn, W.~R.}} \yr{2006}  \at{Upstream nozzle shaping effects on near
  field flow in round turbulent free jets}.  \jt{European Journal of Mechanics
  - B/Fluids}  \bvol{25}~(3),  \pg{279--301}.

\bibitem[Richards \& Pitts(1993)]{Richards1993JoFM417}
{\sc \au{Richards, C.~D.} \& \au{Pitts, W.~M.}} \yr{1993}  \at{Global density
  effects on the self-preservation behaviour of turbulent free jets}.
  \jt{Journal of Fluid Mechanics}  \bvol{254},  \pg{417–435}.

\bibitem[Rodi(1982)]{Rodi1982}
{\sc \au{Rodi, W.}} \yr{1982} {\em Turbulent buoyant jets and plumes\/}, 1st
  edn.  \publ{Oxford; New York: Pergamon Press}.

\bibitem[Schefer {\em et~al.\/}(2008{\natexlab{{\em a\/}}})Schefer, Houf \&
  Williams]{Schefer2008IJoHE4702}
{\sc \au{Schefer, R.~W.}, \au{Houf, W.~G.} \& \au{Williams, T.~C.}}
  \yr{2008{\natexlab{{\em a\/}}}}  \at{Investigation of small-scale unintended
  releases of hydrogen: Buoyancy effects}.  \jt{International Journal of
  Hydrogen Energy}  \bvol{33}~(17),  \pg{4702--4712}.

\bibitem[Schefer {\em et~al.\/}(2008{\natexlab{{\em b\/}}})Schefer, Houf \&
  Williams]{Schefer2008IJoHE6373}
{\sc \au{Schefer, R.~W.}, \au{Houf, W.~G.} \& \au{Williams, T.~C.}}
  \yr{2008{\natexlab{{\em b\/}}}}  \at{Investigation of small-scale unintended
  releases of hydrogen: momentum-dominated regime}.  \jt{International Journal
  of Hydrogen Energy}  \bvol{33}~(21),  \pg{6373--6384}.

\bibitem[Sciacchitano \& Wieneke(2016)]{Sciacchitano2016MSaT84006}
{\sc \au{Sciacchitano, A.} \& \au{Wieneke, B.}} \yr{2016}  \at{Piv uncertainty
  propagation}.  \jt{Measurement Science and Technology}  \bvol{27}~(8),
  \pg{084006}.

\bibitem[Soleimani~nia {\em et~al.\/}(2017)Soleimani~nia, Maxwell, Oshkai \&
  Djilali]{Soleimaninia2017}
{\sc \au{Soleimani~nia, M.}, \au{Maxwell, B.}, \au{Oshkai, P.} \& \au{Djilali,
  N.}} \yr{2017} Measurements of flow velocity and scalar concentration in
  turbulent multi-component jets.  \bt{In {\em Proceedings of the 7th
  International Conference on Hydrogen Safety, Hamburg, Germany\/}}.

\bibitem[{Soleimani nia} {\em et~al.\/}(2018){Soleimani nia}, {Maxwell},
  {Oshkai} \& {Djilali}]{Soleimaninia2018IJoHE}
{\sc \au{{Soleimani nia}, M.}, \au{{Maxwell}, B.}, \au{{Oshkai}, P.} \&
  \au{{Djilali}, N.}} \yr{2018}  \at{{Experimental and Numerical Investigation
  of Turbulent Jets Issuing Through a Realistic Pipeline Geometry: Asymmetry
  Effects for Air, Helium, and Hydrogen}}.  \jt{International Journal of
  Hydrogen Energy}  \bvol{43}~(19),  \pg{9379--9398}.

\bibitem[Su \& Clemens(2003)]{Su2003JoFM1}
{\sc \au{Su, L.~K.} \& \au{Clemens, N.~T.}} \yr{2003}  \at{The structure of
  fine-scale scalar mixing in gas-phase planar turbulent jets}.  \jt{Journal of
  Fluid Mechanics}  \bvol{488},  \pg{1--29}.

\bibitem[Su {\em et~al.\/}(2010)Su, Helmer \& Brownell]{L.K.2010JoFM59}
{\sc \au{Su, L.~K.}, \au{Helmer, D.~B.} \& \au{Brownell, C.~J.}} \yr{2010}
  \at{Quantitative planar imaging of turbulent buoyant jet mixing}.
  \jt{Journal of Fluid Mechanics}  \bvol{643},  \pg{59--95}.

\bibitem[Talbot {\em et~al.\/}(2009)Talbot, Mazellier, Renou, Danaila \&
  Boukhalfa]{Talbot2009EiF769}
{\sc \au{Talbot, B.}, \au{Mazellier, N.}, \au{Renou, B.}, \au{Danaila, L.} \&
  \au{Boukhalfa, M.}} \yr{2009}  \at{Time-resolved velocity and concentration
  measurements in variable-viscosity turbulent jet flow}.  \jt{Experiments in
  Fluids}  \bvol{47}~(4-5),  \pg{769--787}.

\bibitem[Thring \& Newby(1953)]{Thring1953SIoC789}
{\sc \au{Thring, M.~W.} \& \au{Newby, M.~P.}} \yr{1953}  \at{Combustion length
  of enclosed turbulent jet flames}.  \jt{Symposium (International) on
  Combustion}  \bvol{4}~(1),  \pg{789 -- 796}, fourth Symposium (International)
  on Combustion.

\bibitem[Veser {\em et~al.\/}(2011)Veser, Kuznetsov, Fast, Friedrich,
  Kotchourko, Stern, Schwall \& Breitung]{Veser2011IJoHE2351}
{\sc \au{Veser, A.}, \au{Kuznetsov, M.}, \au{Fast, G.}, \au{Friedrich, A.},
  \au{Kotchourko, N.}, \au{Stern, G.}, \au{Schwall, M.} \& \au{Breitung, W.}}
  \yr{2011}  \at{The structure and flame propagation regimes in turbulent
  hydrogen jets}.  \jt{International Journal of Hydrogen Energy}
  \bvol{36}~(3),  \pg{2351--2359}.

\bibitem[{Witze}(1974)]{Witze1974AJ417}
{\sc \au{{Witze}, P.~O.}} \yr{1974}  \at{{Centerline Velocity Decay of
  Compressible Free Jets}}.  \jt{AIAA Journal}  \bvol{12},  \pg{417--418}.

\bibitem[Xiao {\em et~al.\/}(2011)Xiao, Travis \& Breitung]{Xiao2011IJoHE2545}
{\sc \au{Xiao, J.}, \au{Travis, J.R.} \& \au{Breitung, W.}} \yr{2011}
  \at{Hydrogen release from a high pressure gaseous hydrogen reservoir in case
  of a small leak}.  \jt{International Journal of Hydrogen Energy}
  \bvol{36}~(3),  \pg{2545 -- 2554}, the Third Annual International Conference
  on Hydrogen Safety.

\bibitem[Xu \& Antonia(2002)]{Xu2002EiF677}
{\sc \au{Xu, G.} \& \au{Antonia, R.}} \yr{2002}  \at{Effect of different
  initial conditions on a turbulent round free jet}.  \jt{Experiments in
  Fluids}  \bvol{33}~(5),  \pg{677--683}.

\end{thebibliography}

\end{document}